\documentclass[a4paper,11pt]{article}
\pdfoutput=1
\usepackage{jcappub} % for details on the use of the package, please
\usepackage{graphicx}
\usepackage[T1]{fontenc} % if needed
\usepackage{natbib,color,amsmath,amssymb,bm,multirow}
\newcommand{\dd}[1]{\mathrm{d}#1}

\title{\boldmath Reducing Noise in Cosmological N-body Simulations with Neutrinos}

\author[a,b,c]{Arka Banerjee,\note{Corresponding author.}}
\author[a,b,c]{Devon Powell,}
\author[a,b,c]{Tom Abel,}
\author[d]{Francisco Villaescusa-Navarro}
\affiliation[a]{Kavli Institute for Particle Astrophysics and Cosmology, Stanford University, 452 Lomita Mall, Stanford, CA 94305, USA}
\affiliation[b]{Department of Physics, Stanford University, 382 Via Pueblo Mall, Stanford, CA 94305, USA}
\affiliation[c]{SLAC National Accelerator Laboratory, 2575 Sand Hill Road, Menlo Park, CA  94025, USA}
\affiliation[d]{Center for Computational Astrophysics, Flatiron Institute, 162 5th Avenue, 10010, New York, NY, USA}

\emailAdd{arkab@stanford.edu}
\emailAdd{dmpowel1@stanford.edu}
\emailAdd{tabel@stanford.edu}
\emailAdd{fvillaescusa@flatironinstitute.org}

\abstract{We present a new method for generating initial conditions for numerical cosmological simulations in which massive neutrinos are treated as an extra set of N-body (collisionless) particles. It allows us to accurately follow the density field for both Cold Dark Matter (CDM) and neutrinos at both high and low redshifts. At high redshifts, the new method is able to reduce the shot noise in the neutrino power spectrum by a factor of more than $10^7$ compared to previous methods, where the power spectrum was dominated by shot noise at all scales. We find that our new approach also helps to reduce the noise on the total matter power spectrum on large scales, whereas on small scales the results agree with previous simulations. Our new method also allows for a systematic study of clustering of the low velocity tail of the  distribution function of neutrinos. This method also allows for the study of the evolution of the overall velocity distribution as a function of the environment determined by the CDM field.}
\begin{document}
\maketitle
\flushbottom

\section{Introduction}
\label{sec:intro}

The determination of the exact mass scale of the Standard Model neutrinos is one of the most exciting questions for current research in both particle physics and cosmology. Neutrino oscillation experiments have established the mass square differences between the mass eigenstates \cite{SuperK98,SNO2001,K2K2003,DayaBay}, but are not suited for determining either the absolute mass scale of each eigenstate, or the total mass of the three mass eigenstates combined. The high number density of relic thermal neutrinos from the early Universe (e.g. \cite{Dodelson2003book}), and consequently the non-negligible energy fraction in the neutrinos means that various cosmological observables can be used to constrain the masses of neutrinos.

The most immediate effect of the presence of massive neutrinos is to damp the power spectrum on scales below their free streaming scale, relative to the massless case \cite{Hu1997,Lesgourgues2006}. The amount of damping of the power spectrum is proportional to the total energy density in neutrinos, and therefore the total mass of all the neutrino species. However, the exact shape and the scale at which the damping sets in, are, in principle, sensitive to the individual mass of each eigenstate. Various authors have set bounds on the sum of neutrino masses by looking at, for example, the lensing of the CMB \cite{Planck2013,Planck2015,Sherwin2016,deHaan2016}, the Lyman alpha forest power spectrum \cite{Palanque-Delabrouille:2015pga,Palanque-Delabrouille:2014jca}, combining the autocorrelation of galaxies with galaxy lensing and cosmic shear \cite{Abbott:2017wau,Troxel:2017xyo}, and measurements of Baryon Acoustic Oscillations \cite{Giusarma2016,Vagnozzi:2017ovm,Beutler:2014yhv}. Almost all of these results are based on measuring the small scale damping of the power spectrum. While the effect of neutrinos on the power spectrum is well-understood on scales which are linear, i.e. large scales at low redshifts and progressively smaller scales at higher redshifts, fully describing the effects on small scales requires running cosmological simulations which correctly include the effect of massive neutrinos. Since the useful information in various cosmological surveys is proportional to the number of independent modes, and pushing to smaller scales dramatically increases the number of independent modes in the survey, such studies are extremely useful in tightening the bounds on various cosmological parameters, including the neutrino mass. However, these attempts are complicated by the fact that neutrinos, especially at early times, have very high thermal velocities (see e.g., \cite{Lesgourgues2006}), unlike Cold Dark Matter (CDM).

Various attempts have been made to incorporate the effect of massive neutrinos into the standard N-body simulations used to study structure formation in the $\Lambda$-CDM cosmologies. These can be broadly classified into two approaches. The first approach is to use a linear, or perturbative approach for the neutrinos, but coupling to the non-linear gravitational potential sourced by the Cold Dark Matter component \cite{Brandbyge:2008js,Archidiacono:2015ota,Upadhye:2015lia,AliHaimoud:2012vj,Shoji:2010hm,Inman:2016qmg,Senatore:2017hyk,Saito:2008bp,Dakin:2017idt,Wright:2017dkw}. These approaches are useful at intermediate scales and relatively high redshifts, but break down at late times and small scales, depending on the mass of the individual neutrino species being simulated. 

The second approach has been to include neutrinos as an extra set of N-body particles in cosmological simulations \cite{Viel:2010bn,Bird:2011rb,Brandbyge:2009ce,Villaescusa-Navarro:2013pva,Costanzi:2013bha,Castorina:2013wga,Castorina2015,Carbone:2016nzj,Yu:2016yfe,Emberson:2016ecv,RSD2017,Adamek:2017uiq}. In addition to a bulk velocity determined by the power spectrum, each neutrino particle is also given a random thermal velocity by sampling the Fermi-Dirac distribution. This method is fully non-linear, but suffers from Poissonian shot-noise.
%the traditional way of generating the initial conditions means that the neutrino field in these simulations is shot-noise dominated, especially at early times. 
The reason is that neutrinos have very large thermal velocities at high-redshifts, which allows them to cross the simulation box multiple times. Because of this, neutrino particles quickly lose memory about their initial conditions and distribute themselves randomly in the box, giving rise to shot-noise in their power spectrum. N-body methods rely on the assumption that the number of particles in a given volume is a faithful representation of the actual physical density in that volume. However, for neutrinos free-streaming with large thermal velocities, this assumption is no longer valid, and the number of neutrino particles in a given volume is a just described by Poissonian statistics. This also means that error on the neutrino power spectrum only improves as $1/N$ with the number of particles used in these simulations. To completely remove shot noise from typical simulations with massive neutrinos, just by increasing the number of particles, one would need roughly need a factor of $10^7$ more particles than used in the largest simulations today! While some methods have been suggested to get rid of the noise in the neutrino power spectrum by sub-sampling the particles to get two independent realizations of the density field (see e.g. \cite{Adamek:2017uiq,Inman:2015pfa}), it does not actually get rid of the inherent noise in field that is used to source the Poisson equation. Therefore new methods are required to test the accuracy of various quantities measured from simulations run using this technique. This is especially true in the context of upcoming cosmological surveys which can potentially pin down the matter power spectrum to the accuracy of $1\%$ \cite{DESI,EUCLID,WFIRST,LSST} at a range of scales.

While \cite{Banerjee2016} has proposed a method to remove the shot noise issue while fully capturing the non-linearities, using a combination of N-body and hydrodynamic techniques, in this paper we describe a completely new approach to remove shot noise from simulations of massive neutrinos, run using the standard N-body technique, by just changing the initial conditions used to initialize these simulations. 

The plan of the paper is as follows: we describe our new method for generating initial conditions in \S \ref{sec:method}. In \S \ref{sec:sims}, we describe the details of the cosmological simulations that are used in the paper. In \S \ref{sec:results}, we present results at high redshifts and at $z=0$ to explore the convergence properties and accuracy of our new scheme. We also compare our results to those obtained from simulations with shot noise, as well as various fitting functions which try to capture the effect of massive neutrinos on the matter power spectrum. In \S \ref{sec:sheets}, we explore how this method also allows us to trace shells of different velocities in the neutrino phase space, and allow for a better understanding of the structure of the distribution function of neutrinos in different environments, such as halos and voids. Finally in \S \ref{sec:disc}, we summarize our results, and discuss various future directions of work.

\section{Method}
\label{sec:method}
Before decoupling, neutrinos follow a thermal Fermi-Dirac distribution. Once the neutrinos decouple from the rest of the universe, the distribution gets frozen in, but the momenta of the neutrinos redshift as $1/a$, where $a$ is the cosmological scale factor. As long as the neutrinos were relativistic at the time of decoupling, this is equivalent to having a Fermi-Dirac distribution with the temperature falling as $T\propto 1/a$. Therefore the neutrino distribution at a redshift $z$ is given by
\begin{equation}
\label{eq:fd}
f(\mathbf p) = \frac{4\pi g_\nu}{(2\pi \hbar)^3} \frac 1 {e ^{\frac{pc}{k_{\footnotesize B} T_\nu (1+z)}}+1} \, ,
\end{equation}
where $p = \sqrt{\mathbf p . \mathbf p}$, and $g_\nu$ is the degeneracy (three neutrino species and three anti-neutrino species). $T_\nu(z) \simeq 1.95\,$K is the neutrino temperature today. Our aim is to sample the velocity or momentum space in a regular manner, which is replicated at every point in configuration space - i.e. at every point on the grid used to generate the initial conditions. This sort of approach has been adopted previously in a different context in \cite{1970PhFl...13.1819B}. An inspection of Eq. \ref{eq:fd} shows that choosing to tile the velocity space using a Cartesian grid structure is difficult as the Fermi-Dirac distribution does not factorize along the three directions. On the other hand, the distribution looks much simpler if one considers the magnitude of the momentum, the distribution for which is given simply by $p^2\, f(p)$. Therefore, we choose to first decompose the distribution into intervals of the magnitude of velocity, and then choose angular directions for each velocity magnitude. We discuss the method for each step in detail below. 

The reason this method is effective at removing shot noise is that it fixes the magnitudes and directions of the thermal velocities of the neutrino particles. This means that the number of neutrino particles moving into and out of some given volume of the simulation box is not a random number, as is the case when the distribution is sampled randomly. Instead, it is completely determined by the physical power spectrum of the neutrinos, which sets the initial displacements off grid points, and the gravitational evolution of the trajectories. For example, if we decided to initialize the neutrino particles exactly off grid points, i.e. without a physical power spectrum or a completely uniform initial density field, the regular sampling of magnitudes and directions would mean that, in the absence of gravity, at any later time, the density field would still be exactly uniform. Obviously, were we to choose velocity magnitudes and directions randomly, this would not be true, even if we initialized the neutrino particles exactly off the grid points, and gravity is turned off. Note that the ``pairing'' scheme (see for e.g. \cite{Ma:1993xs}), where after the velocity magnitude and direction are chosen at random, two neutrino particles are generated in opposite directions to conserve momentum, would also suffer from the same problem. This is because the neutrino particles generated at adjacent grid points would still have different magnitudes and directions for the velocity. 

We now discuss the recipe for dividing the Fermi-Dirac distribution into shells of equal ``mass" or phase space volume. In order to divide the distribution into $N_{\rm shell}$ shells, we construct each shell $i$ such that 
\begin{equation}
\label{shell_division}
\frac{\int_{p_{min}^i}^{p_{max}^i} p^2 f(p) \, \dd{p}}{\int_{0}^{p_{max}} p^2 f(p) \, \dd{p}} = \frac 1 {N_{\rm shell}} \, ,
\end{equation}
where $p_{min}^0 = 0$, $p_{min}^i$ and $p_{max}^i$ are the minimum and maximum value of momentum for each bin, and $p_{max}$ is some reasonable value where we can truncate the distribution. In practice, we work in the dimensionless units of $p/T$, which has the additional advantage of making $p_{max}$ independent of the neutrino temperature. We set the magnitude of the velocity associated with each shell in the following manner:
\begin{equation}
\label{eq:shell_vel}
\langle p_i \rangle = \sqrt{\frac{\int_{p_{min}^i}^{p_{max}^i} p^4 f(p) \, \dd{p}}{\int_{p_{min}^i}^{p_{max}^i} p^2 f(p) \, \dd{p}}} \, .
\end{equation}
Eq. \ref{eq:shell_vel} ensures that the velocity dispersion of the neutrinos is set to the correct analytic value. 

As we will discuss in detail in \S3, we also need a recipe for dividing up the Fermi-Dirac distribution in a way that allows us to better resolve the slower moving neutrino shells. We do this in the following manner. We choose a weighting function $g(p)$ such that $g(p)$ has higher weight at low $p$, and lower weight at high $p$, compared to $p^2\, f(p)$. For example, in our calculations, we have used $g(p) = p \, f(p)$, and $g(p) = \ln(1+p) f(p)$. Once the function $g(p)$ has been chosen, the different shells are defined in terms of their $p_{\rm min}^i $ and $p_{\rm max}^i$ similar to Eq.\ref{shell_division},
\begin{equation}
\label{shell_division_unequal}
\frac{\int_{p_{min}^i}^{p_{max}^i} g(p) \, \dd{p}}{\int_{0}^{p_{max}} g(p) \, \dd{p}} = \frac 1 {N_{\rm shell}} \, .
\end{equation}
The fractional mass in each shell is calculated in the following manner:
\begin{equation}
\label{shell_mass}
m^i_{\rm shell} = \frac{\int_{p_{min}^i}^{p_{max}^i} p^2 f(p) \, \dd{p}}{\int_{0}^{p_{max}} p^2 f(p) \, \dd{p}} \, .
\end{equation}
So, for this method neutrino particles from different shells have different masses. We can still use Eq. \ref{eq:shell_vel} to determine the magnitude of the velocity corresponding to each shell.   

\begin{figure}[tbp]
\centering 
\includegraphics[width=.7\textwidth]{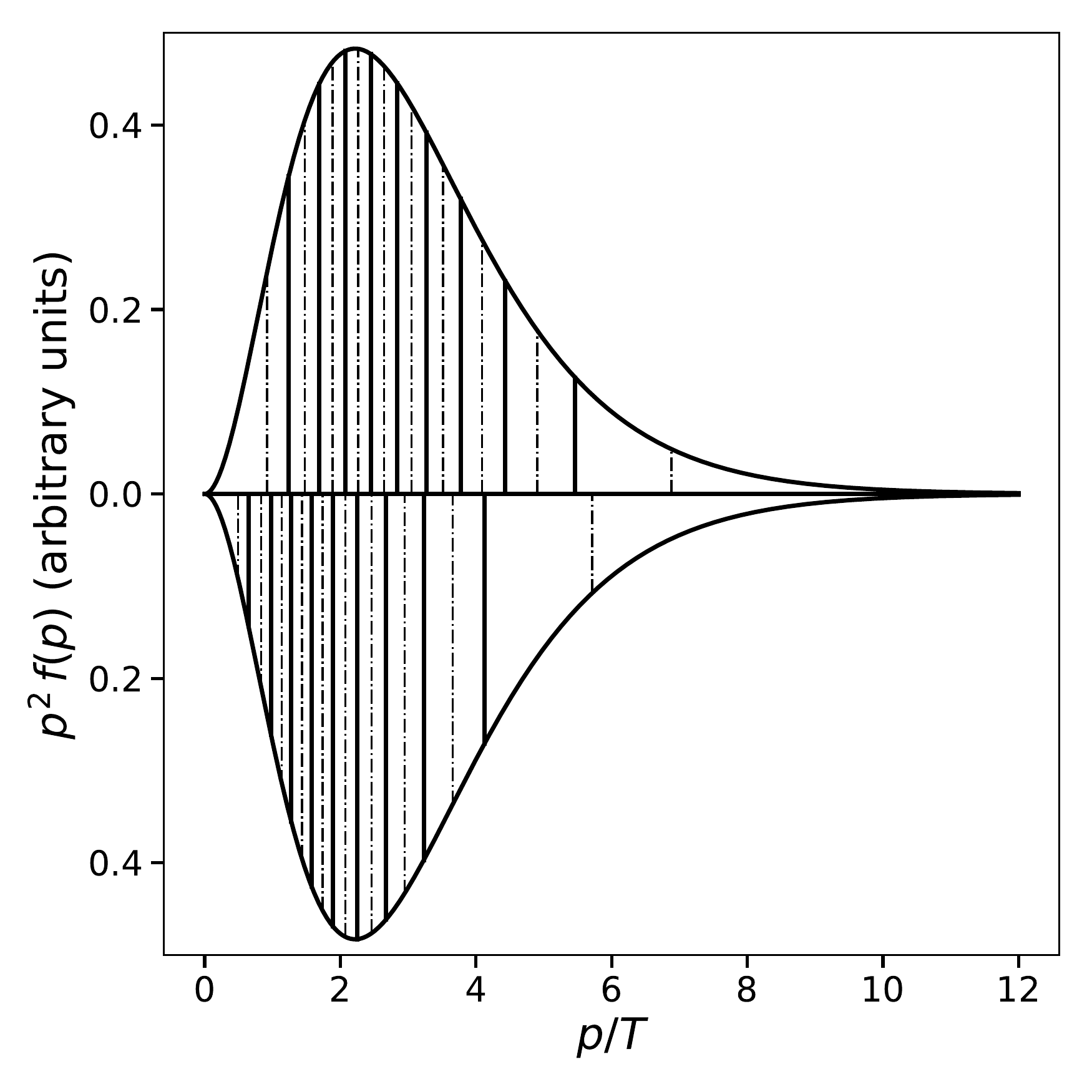}
\caption{\label{fig:l} Illustration of our method for dividing up the initial Fermi-Dirac distribution for neutrinos. The solid lines represent the boundaries of different momentum bins, while the dashed lines represent the velocity associated with each shell ($\langle p_i\rangle$). On the top, we show the case when all the subdivisions have the same mass or phase-space volume. This has more resolution near the peak of the distribution. At the bottom, we show the subdivisions for the case where $g(p)$ as defined in the text is taken to be $g(p) = p \, f(p)$. This puts more resolution on the slower end of the distribution function. Each sub-division of the distribution corresponds to a ``shell'' in our terminology.}
\end{figure}

Once we have divided up the distribution function into shells in velocity space, we subdivide each of these spherical shells into equal area elements using the \texttt{HEALPIX} algorithm ~\cite{healpix}. At the lowest refinement level, \texttt{HEALPIX} tiles the surface of the sphere with 12 elements. Higher refinement levels are labeled by the variable $N_{\rm side}$, where for a given value of $N_{\rm side}$, the sphere is divided into $12 N_{\rm side}^2$ elements. \texttt{HEALPIX} elements are always symmetric, irrespective of the value of $N_{\rm side}$. However, for low $N_{\rm side}$, especially $N_{\rm side} = 1$ and $N_{\rm side} =2$, there are some residual anisotropies. This comes from the fact that the equatorial tiling and polar tilings are very different, causing the $\hat z$ direction to be different from the $\hat x$ and $\hat y$ directions. For low $N_{\rm side}$, we try to minimize the effects of this anisotropy by rescaling the velocities to ensure that the dispersion along each of the three directions is the same. In summary, therefore, the magnitude of the velocity is chosen by dividing up the Fermi-Dirac distribution; the direction of the velocity is given by the \texttt{HEALPIX} algorithm, and this procedure is repeated for every point on the grid on which the initial conditions are generated. While we have used \texttt{HEALPIX} to generate the velocity directions, other schemes of uniformly dividing the unit sphere \cite{2016arXiv160704590H} would also produce similar effects as those discussed in this work.

\section{Simulations}
\label{sec:sims}

To generate cosmological initial conditions with this method, we modify a version of N-GenIC, that computes displacements and peculiar velocities accountting for the fact that in cosmologies with massive neutrinos the growth factor and the growth rate are scale-dependent \cite[e.g.][]{RSD2017, Raccanelli_2017, Roncarelli_2017, Zennaro:2016nqo, Massara_2015}. The simulations are then run with the Gadget-3 cosmological N-body code, which is a modified version of the publicly available Gadget-2 code \cite{Springel:2005mi}. We note that we turn off the short-range force for neutrinos at early times - turning it on only at redshift $z=9$. The reason we do this at early times is that our method for generating initial conditions produces multiple neutrino particles at the same position. While constructing the tree in Gadget, particles at very close positions are randomized in order to complete the tree construction, leading to artifacts in the simulation. We check that our results are not sensitive to the exact redshift at which we turn on the tree construction for the neutrinos.

We run all our simulations for a comoving box size of $1$ Gpc/h. The cosmological parameters we use are the following: $\Omega_m = 0.3175$, $\Omega_b = 0.049$, $\Omega_\Lambda = 0.6825$, $n_s = 0.9624$, and $h = 0.6711$. Note that in our parameterization $\Omega_\nu$ is always included in $\Omega_m$, i.e. $\Omega_{\rm m}=\Omega_{\rm cdm}+\Omega_{\rm b}+\Omega_\nu$.

In order to generate the initial conditions for high redshift studies in \S \ref{sec:conv}, each point on the initial grid starts $12\times N_{\rm side}^2\times N_{\rm shell}$ neutrino particles, along with one CDM particle. For these studies, all initial conditions were generated with equal mass neutrinos. The particles are displaced off the grids using first order Lagrangian perturbation theory (Zeldovich approximation). The masses of the neutrino particles are adjusted so that they add up to give the correct $\Omega_\nu$ for the simulation box. Of course, this implementation is very expensive in terms of particle numbers, and can only be used to run convergence tests with small numbers of CDM particles. These runs correspond to the first five simulations shown in Table \ref{tab:a}.

For simulations which run to $z=0$, in \S \ref{sec:full_runs}, we follow a different strategy. We generate the initial conditions for the CDM particles from a fine grid, and for the neutrinos from a coarser grid. Therefore, we will have $N_{\rm fine}^3$ CDM particles, and $12\times N_{\rm side}^2 \times N_{\rm shell} \times N_{\rm coarse}^3$ neutrino particles. By playing around with the sizes of the two grids, we can ensure that the total number of CDM particles and neutrino particles are of the same magnitude, as is the case in most simulations run with CDM and neutrino particles. We list the choices of parameters for various runs in Table \ref{tab:a}. For the simulations with massive neutrinos represented in Table \ref{tab:a}, we consider the degenerate neutrino mass scenario, where all the individual mass eigenstates are equal in mass. We do consider one specific case where there is only one massive neutrino species - this case is discussed in \S \ref{sec:results}.

\begin{table}
\begin{center}
\begin{tabular}{| c | c | c | c | c | c | c | c |}
\hline
Name & $\sum m_\nu$ & $N_{\rm CDM}^{1/3}$ & RV & $N_{\rm coarse}$ & $N_{\rm shell}$ & $N_{\rm side}$ & EM \\
\hline
\texttt{E\_Rand\_15} & 0.15eV & 128 & Y & - & - & - & Y \\
\hline
\texttt{E\_Rand2\_15} & 0.15eV & 128 & Y & - & - & - & Y \\
\hline
\texttt{E\_Rand3\_15} & 0.15eV & 128 & Y & - & - & - & Y \\
\hline
\texttt{E\_SH10\_NS1} & 0.15eV & 128 & N & 128 & 10 & 1 & Y \\
\hline
\texttt{E\_SH40\_NS1} & 0.15eV & 128 & N & 128 & 40 & 1 & Y \\
\hline
\texttt{E\_SH10\_NS2} & 0.15eV & 128 & N & 128 & 10 & 2 & Y \\
\hline
\texttt{E\_SH10\_NS4} & 0.15eV & 128 & N & 128 & 10 & 4 & Y \\
\hline
\texttt{E\_SH5\_NS4} & 0.15eV & 128 & N & 128 & 5 & 4 & Y \\
\hline
\texttt{L\_CDM} & 0 eV & 512 & - & - & - & - & - \\
\hline
\texttt{L\_Rand\_15} & 0.15eV & 512 & Y & - & - & - & Y \\
\hline
\texttt{L\_Rand\_30} & 0.30eV & 512 & Y & - & - & - & Y \\
\hline
\texttt{L\_SH5\_NS1} & 0.15eV & 512 & N & 256 & 5 & 1 & Y \\
\hline
\texttt{L\_SH5\_NS2} & 0.15eV & 512 & N & 128 & 5 & 2 & Y \\
\hline
\texttt{L\_SH5\_NS4} & 0.15eV & 512 & N & 128 & 5 & 4 & Y \\
\hline
\texttt{L\_SH10\_NS2} & 0.15eV & 512 & N & 128 & 10 & 2 & Y \\
\hline
\texttt{LU\_SH10\_NS2} & 0.15eV & 512 & N & 128 & 10 & 2 & N (scheme 1) \\
\hline
\texttt{LU2\_SH10\_NS2} & 0.15eV & 512 & N & 128 & 10 & 2 & N (scheme 2) \\
\hline
\texttt{LU\_SH20\_NS2} & 0.15eV & 512 & N & 128 & 20 & 2 & N (scheme 1)\\
\hline
\texttt{LU3\_SH10\_NS2} & 0.30eV & 512 & N & 128 & 10 & 2 & N (scheme 1)\\
\hline
\end{tabular}
\caption{\label{tab:a} A summary of the parameters of various runs. The column labeled ``RV" corresponds to whether the neutrino particles were given random thermal velocities. The column labeled ``EM'' indicates if all the neutrino particles in the simulation had the same masses  (Y) or not (N). Scheme 1 corresponds to setting the function $g(p) = p\, f(p)$, while scheme 2 corresponds to setting $g(p) = \ln(1+p) \, f(p)$. Note that the run \texttt{E\_Rand\_15} used the same number of neutrino particles as \texttt{E\_SH10\_NS1}, while \texttt{E\_Rand2\_15} had the same number of neutrino particles as \texttt{E\_SH10\_NS2}, and \texttt{E\_Rand2\_15} used the same of neutrino particles as \texttt{E\_SH10\_NS4}.}
\end{center}
\end{table}

\section{Results}
\label{sec:results}
\subsection{High redshift}
\label{sec:conv}

The shot noise arising from the thermal motion of neutrino particles in simulations is more dominant at early redshifts. This is because at early redshifts, the amplitude of the neutrino power spectrum is small compared to the $1/N$ shot noise power spectrum on almost all scales in the simulation volume. As the thermal velocities of the neutrinos goes down, due to the Universe's expansion, and the physical perturbations grow, the scale at which the noise power spectrum dominates over the physical power spectrum moves to smaller values. Therefore, to display the effectiveness of our initial conditions to get rid of the shot noise in the neutrino power spectrum is best demonstrated by comparing the results of simulations initialized by our method to those from simulations initialized using earlier methods of sampling the thermal velocities of neutrinos. Also, at high redshifts, linear perturbation theory holds at all scales of interest, and therefore we can safely compare the results of our simulations to linear theory predictions from Boltzmann codes like \texttt{CAMB} \cite{CAMB} and \texttt{CLASS} \cite{CLASS1} without worrying about non-linear effects.

\begin{figure}[tbp]
\centering 
\includegraphics[width=.6\textwidth]{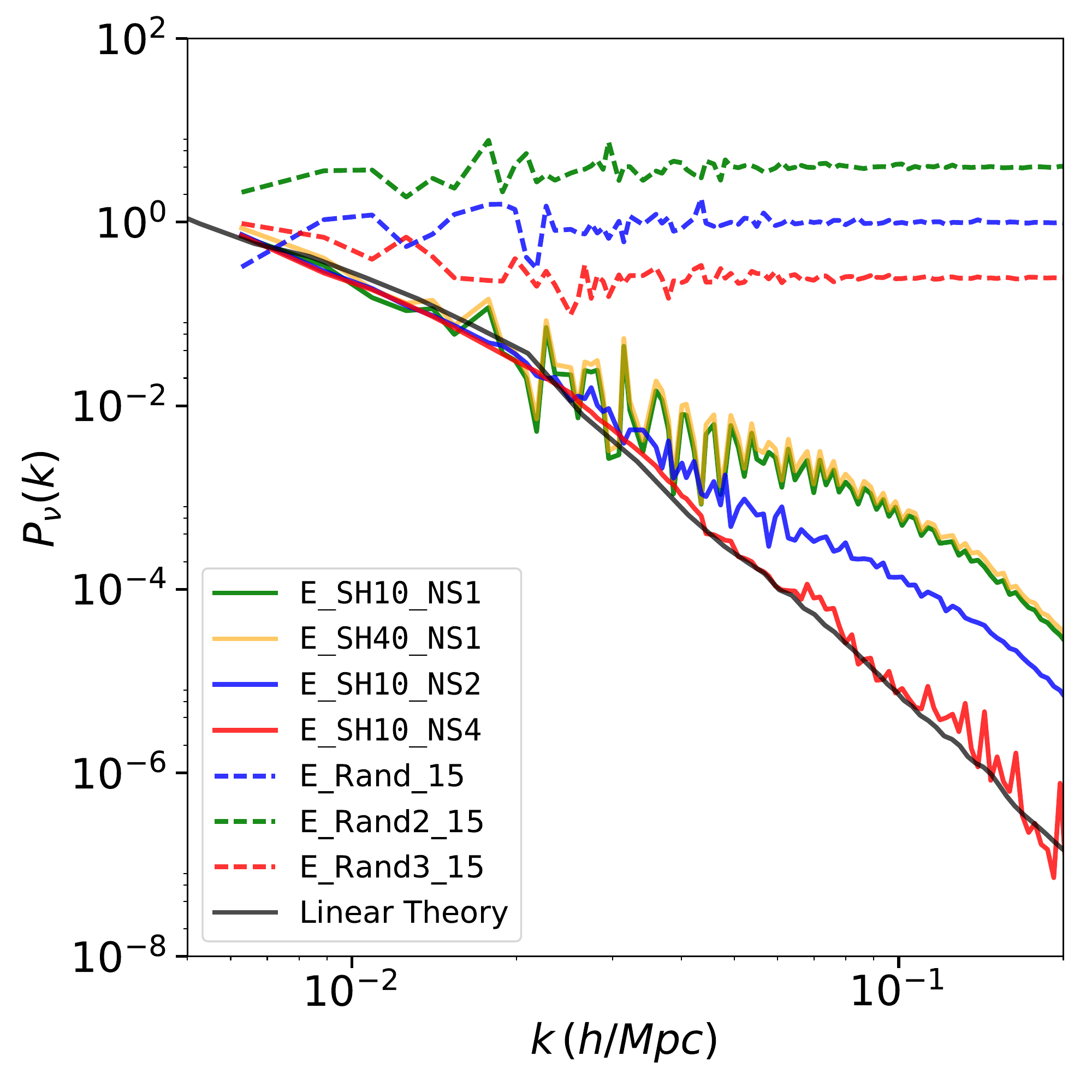}
\caption{\label{fig:c} Comparison of the neutrino power spectrum at $z=49$ for various choices of $N_{\rm shell}$ and $N_{\rm side}$. We compare the results from our simulations (solid curves) to linear theory (black), as well as a simulations where the Fermi Dirac distribution was sampled randomly for the thermal velocities (dashed curves). For each color, the solid curve and dashed curve used the same number of neutrino particles in the simulations. The curves for \texttt{E\_SH10\_NS1} and \texttt{E\_SH40\_NS1} have been displaced slightly from each other to make them visible on the plot. We find that at early times, the isotropy of the neutrino distribution (higher $N_{\rm side}$) is more important for matching the linear theory prediction for the neutrino power spectrum compared to the number of radial shells ($N_{\rm shell}$).}
\end{figure}

To perform the comparisons, we initialize our simulations at $z=99$, and run up to $z=49$. We explore the case where the sum of the neutrino masses $\sum m_\nu = 0.15$eV, and each individual mass eigenstate has mass $m_i = 0.05$eV. For these light neutrinos, the free streaming scale is larger than the simulation box size of $1$ Gpc/h, implying that the neutrino transfer function is damped compared to the CDM transfer function at all scales. 

All of the runs considered in this section were performed with $128^3$ CDM particles, and the same grid size was used to generate the neutrino initial conditions. For a fixed size of the initial conditions grid, there are two parameters which can be changed in our method - the number of radial velocity shells, $N_{\rm shell}$, and the number of \texttt{HEALPIX} pixels, controlled by $N_{\rm side}$. As we increase $N_{\rm shell}$, we sample the radial velocity distribution more finely. Increasing $N_{\rm side}$ results in more more directions in which the neutrinos can move, and results in increasing the isotropy of the distribution. 

As pointed out previously, the method in this paper generates multiple neutrino particles per point of the initial conditions grid. To make a fair comparison to the results of simulations initialized using the recipes followed in previous studies, we modify the procedure so that we use the same number of neutrino particles to randomly sample the Fermi-Dirac distribution as are used in runs with initial conditions generated using the new method. 

We plot the results of our study in Fig. \ref{fig:c}, where the solid curves are from the ``noiseless" simulations and the dashed curves are from simulations with shot noise. Curves of the same color used the same numbers of neutrino particles. We find that even with increased number of particles, the neutrino power spectrum from the randomly sampled runs (dashed curves) is shot-noise dominated at all scales in the simulation box. This is because the physical power spectrum of the neutrinos at this redshift is smaller than the shot noise floor determined by particle number. We also find that for a fixed $N_{\rm side}$, changing $N_{\rm shell}$ does not have a very big effect, as can be seem by comparing the results from runs \texttt{E\_SH10\_NS1} and \texttt{E\_SH40\_NS1}. The first run uses 10 radial shells, while the second uses 40 radial shells, with $N_{\rm side} =1$ for both. On the other hand, if we fix $N_{\rm shell}$ and increase $N_{\rm side}$, we move closer to the linear theory prediction for the neutrino power spectrum, and for $N_{\rm side} = 4$, we obtain a pretty good match with theory. While there is still some noise in the neutrino power spectrum, we are able to beat down shot noise in the power spectrum by a factor of nearly $10^8$.  

These results imply that at early redshifts, ensuring the isotropy of the neutrino distribution (higher $N_{\rm side}$) is more important than sampling the Fermi-Dirac distribution finely (higher $N_{\rm shell}$). This is not very surprising, because at early times, all the radial velocity shells redshift in the same way. It is therefore possible to accurately reconstruct the Fermi-Dirac distribution with a small number of points. This would no longer be true if each velocity shell behaved differently, as will happen at low redshifts, and we explore this behavior in detail in the next section.

\subsection{Low redshift}
\label{sec:full_runs}

We looked at the neutrino power spectrum at high redshifts to demonstrate the ability of our method to eliminate Poisson noise from the simulations. However, this is currently not directly observable, so at low redshifts, we shift our attention to the total matter auto power spectrum, and the neutrino-CDM cross power spectrum. In this section, we perform a study of the two spectra at $z=0$ for various parameter choices. We compare the results from these simulations to simulations which have shot noise in them, as well as to various fitting functions from literature.

To compute the matter power spectrum, we can first define the matter overdensity field in terms of the overdensity fields of the CDM component ($\delta_c$) and the neutrinos ($\delta_\nu$) as 
\begin{equation}
\delta_m = f_c \delta_c + f_\nu \delta_\nu \,
\end{equation}
where $f_c$ and $f_\nu$ are the fractions of the total matter density that is in each component. That is,
\begin{equation}
f_c = \frac{\Omega_c}{\Omega_c + \Omega_\nu} \qquad ; \qquad f_\nu = \frac{\Omega_\nu}{\Omega_c + \Omega_\nu} \, .
\end{equation}
The matter power spectrum is then given by
\begin{equation}
P_{mm} = f_c^2 P_{cc} + 2 f_c f_\nu P_{c\nu} + f_\nu^2 P_{\nu\nu} \,,
\label{eq:Pkmm}
\end{equation}
where $P_{cc}$ and $P_{\nu\nu}$ are the auto-spectra of the CDM and neutrino fields, and $P_{c\nu}$ is the cross spectrum.

For the simulations considered here, we generate the initial conditions for the CDM particles from a $512^3$ grid. As in the previous section, all simulations were initialized at $z=99$. For the neutrinos, we explore a range of parameters allowed by computational resources. We vary both $N_{\rm shell}$ and $N_{\rm side}$, as well as the size of the grid on which the initial conditions for the neutrinos are generated. The last is to study the effects of the Nyquist frequency of the neutrino particles on the different observables. We concentrate on $\sum m_\nu=0.15$eV, with individual masses $m_i = 0.05$eV. We also present some results for the case $\sum m_\nu = 0.3$eV, with $m_i = 0.1$eV.

\begin{figure}[tbp]
\centering 
\includegraphics[width=.475\textwidth]{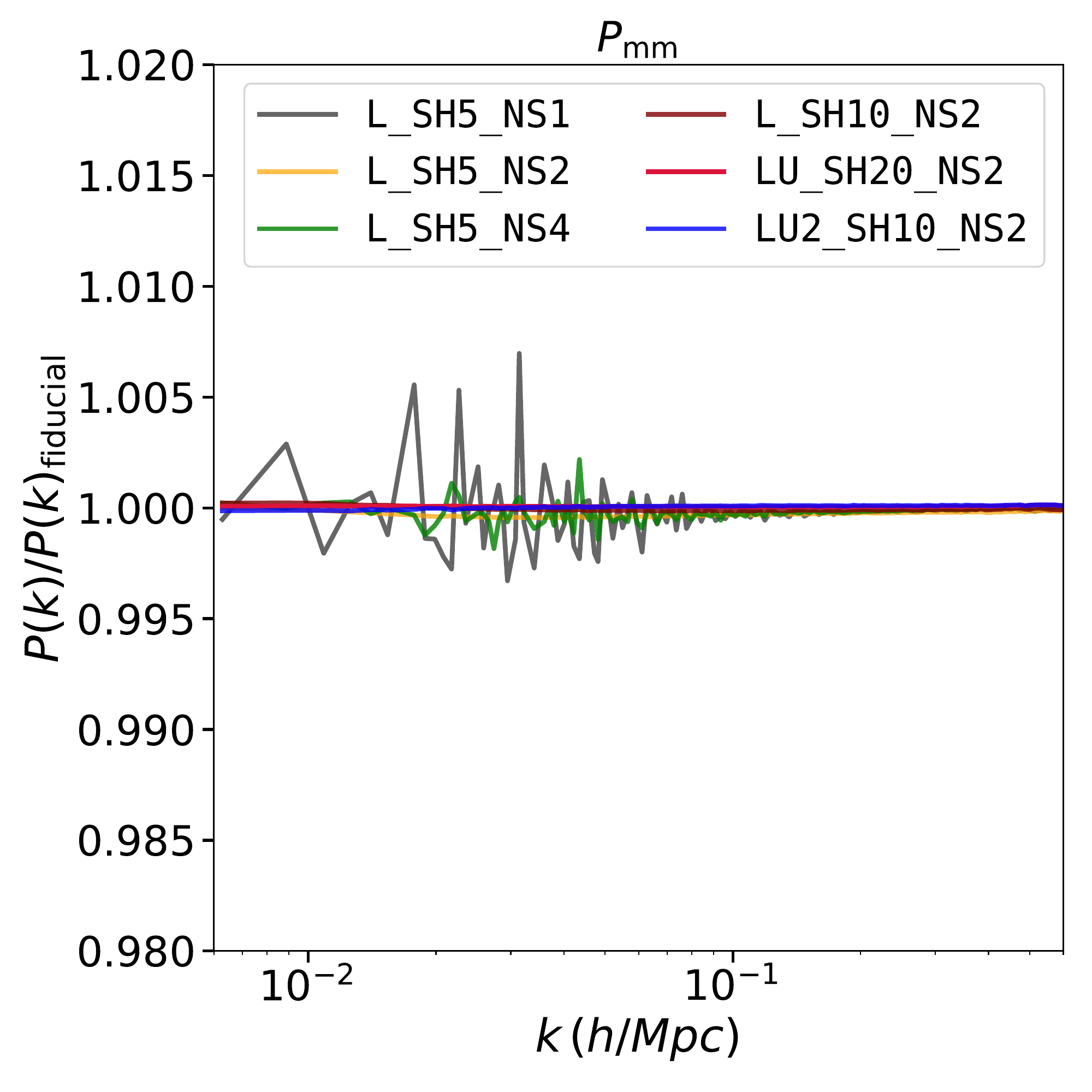}
\includegraphics[width=.475\textwidth]{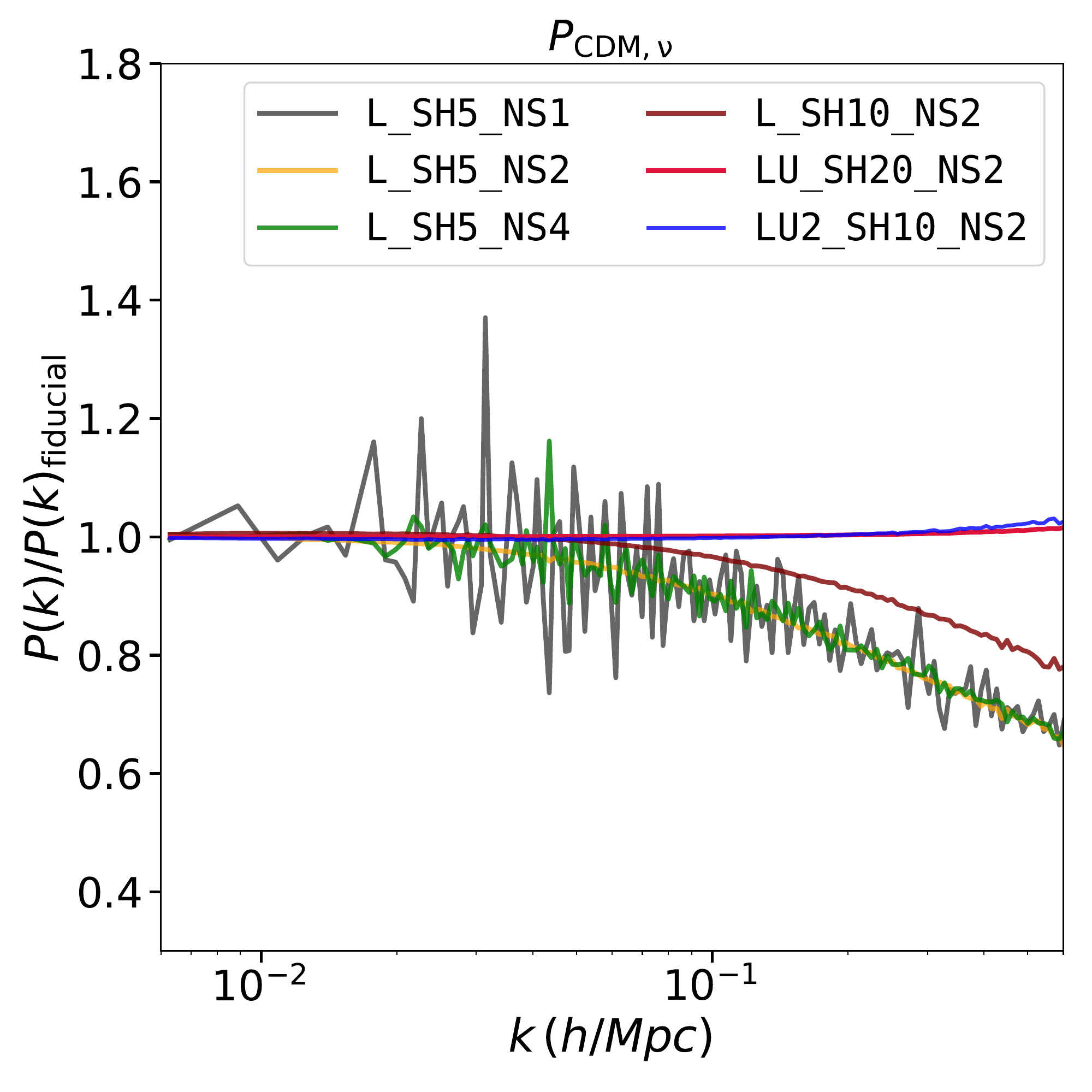}
\caption{\label{fig:f} We look at the convergence of the low redshift results with different number of radial shells into which the Fermi Dirac distribution is divided, and different number of angular directions in which the neutrinos can move initially. The matter power spectrum (left panel) is well converged for most choices of parameters. For the CDM-$\nu$ cross-power spectrum, we find that $N_{\rm shell}$ is important to resolve the full clustering of neutrinos. Having higher resolution on the lower velocity shells helps with the convergence tests. The fiducial simulation used here is \texttt{LU\_SH10\_NS2}.}
\end{figure}

The results of the various runs are summarized in Fig. \ref{fig:f}. We choose the run \texttt{LU\_SH10\_NS2} as our fiducial run. This run has $N_{\rm shell}=10$, $N_{\rm side}=2$, and a $128^3$ grid was used to generate the neutrino initial conditions. This run also uses unequal neutrino particle masses to allow better resolution of the slow-moving shells from the Fermi-Dirac distribution. If we look at the ratio of the matter power spectrum for various runs compared to the fiducial case (the left panel of Fig. \ref{fig:f}), we find that it is remarkably well converged. The run with $N_{\rm side} = 1$, \texttt{L\_SH5\_NS1} is somewhat noisy on large and intermediate scales, but agrees well with the fiducial run on small scales. Since this run (\texttt{L\_SH5\_NS1}) used a $256^3$ grid to generate the neutrino initial conditions, the agreement at larger values of $k$ implies that, at least on the scales we are interested in, increasing the neutrino grid size beyond the fiducial value, does not affect the computation of the matter power spectrum. Increasing $N_{\rm side}$ from $2$ to $4$ produces very little change, with the noise level at about $0.2\%$. 

Similarly, we find that as we increase $N_{\rm shell}$ while keeping fixed the other parameters, the ratio of the matter power spectrum barely changes. In terms of the matter power spectrum alone, therefore, as long as $N_{\rm side}$ is larger than $1$, the result is well-converged on all scales. This suggests that, just like at high redshifts, it is important that the neutrino distribution is isotropic to a certain degree. We also point out that for all the runs, the mean of the ratio is very close to $1$, implying that the matter power spectrum is insensitive to whether we use equal mass neutrino or neutrino particles with unequal masses, as is the case with the fiducial run \texttt{LU\_SH10\_NS2}.
While the matter power spectrum is, of course, the most important observable that is affected by the presence of massive neutrinos, and that can be calculated from these simulations, we also look at the cross-power spectrum between the CDM and the neutrino fields. While this quantity may not be directly observable, it is an important piece to model the total matter distribution and the non-linear coupling between CDM and neutrinos.
%this cross spectrum allows us to study the extent of the neutrino clustering around the cosmic web laid down by the CDM component. 
The results of this investigation is presented in the right panel of Fig. \ref{fig:f}. As with the matter power spectrum, we present the ratio of the cross spectrum normalized by the cross spectrum from the fiducial run \texttt{LU\_SH10\_NS2}. We find that as $N_{\rm side}$ is increased from $1$ to $2$ to $4$, while keeping $N_{\rm shell}$ fixed at 5, the mean behavior does not change. There is some amount of noise about the mean as we change $N_{\rm side}$.  

There is, however, a trend in the mean behavior as we increase $N_{\rm shell}$ for a fixed $N_{\rm side}$. We find that for $k \geq 2 \times 10^{-2} \, h/$Mpc, the magnitude of the cross spectrum increases as we increase the number of radial shells that are used to split up the Fermi-Dirac distribution from $5$ to $10$ - going from \texttt{L\_SH5\_NS2} to \texttt{L\_SH10\_NS2}. While we have not plotted it in the figure to reduce the number of curves, this trend continues as we further increase $N_{\rm shell}$ from $10$ to $20$. This indicates that at late times, the clustering of the neutrinos about the CDM field, as measured by the cross spectrum, is dominated by the slowest moving neutrinos, as expected from theoretical calculations \cite{Massara_2014}, and disproportionate to the phase space ``mass'' in these particles. To clarify, we find this trend in the simulations run with equal neutrino particle masses. For these runs, we divided up the initial Fermi-Dirac distribution such that the different shells covered equal volumes in phase space - i.e. there is more resolution in near the peak of the initial Fermi-Dirac distribution than either at the high-velocity tail or the low-velocity end of the distribution. However, at low redshifts, the slowest moving neutrinos will be the ones whose motion is most affected by the gravitational potential sourced by the CDM field, and they contribute the most to the cross correlation power spectrum that we measure from the simulations.

This motivates us to consider situations where there is more resolution on the low velocity end of the Fermi-Dirac distribution, while also indicating that we do not require much resolution for the higher velocity parts of the distribution. This is exactly what we do for the runs with unequal neutrino particles mass - we use the function $g(p)$ as outlined in \S \ref{sec:method} to ensure that the lower velocity parts of the distribution are populated by larger numbers of particles in the simulations. When we do this, we find that we get convergence in the cross correlation spectrum by using $N_{\rm shell} = 10$ (fiducial run), as the result does not change much when we increase $N_{\rm shell}$ to $20$ (\texttt{LU\_SH20\_NS2}). We also find that the result is not very sensitive to our choice of the function $g(p)$ between $g(p) = p \, f(p)$ and $g(p) = \ln (1+p) \, f(p)$, as seen by inspecting the curve from the simulation \texttt{LU2\_SH10\_NS2}, where the latter choice was used to generate the neutrino particle masses and velocities. The fiducial run used the first choice for $g(p)$ to generate masses and velocities. We further explore the clustering of different parts of the initial Fermi-Dirac distribution of velocities in \S \ref{sec:sheets}.

We have thus demonstrated the convergence of the results of our simulations, in terms of the matter power spectrum and the CDM-neutrino cross-power spectrum, as a function of the free parameters of our method for generating initial conditions. As the next step, we compare the results from these simulations to those produced by the same N-body method, but with previous methods of generating initial conditions. As discussed previously, the older method samples the thermal velocity distribution randomly, which leads directly to the shot noise in the neutrino density field. Comparing the results from the two methods, is therefore a direct way of testing the effects of the shot noise on different observables. To clarify, we will compare the results of the fiducial simulations \texttt{LU\_SH10\_NS2}, and \texttt{LU3\_SH10\_NS2}, to the runs with random velocities. We also point out that the simulations with random velocities will, in general, have lower numbers of neutrino particles compared to the runs mentioned above, since these runs are usually initialized with $1$ neutrino particle per grid point on the initial conditions grid. We also compare two different scenarios - one where $\sum m_\nu = 0.15\,$eV, and the other where $\sum m_\nu = 0.3\,$eV. This helps us understand how the simulations compare as $f_\nu$ is varied. In both cases, we focus on the degenerate scenario, where each mass eigenstate has the same mass. 

\begin{figure}[tbp]
\centering 
\includegraphics[width=.45\textwidth]{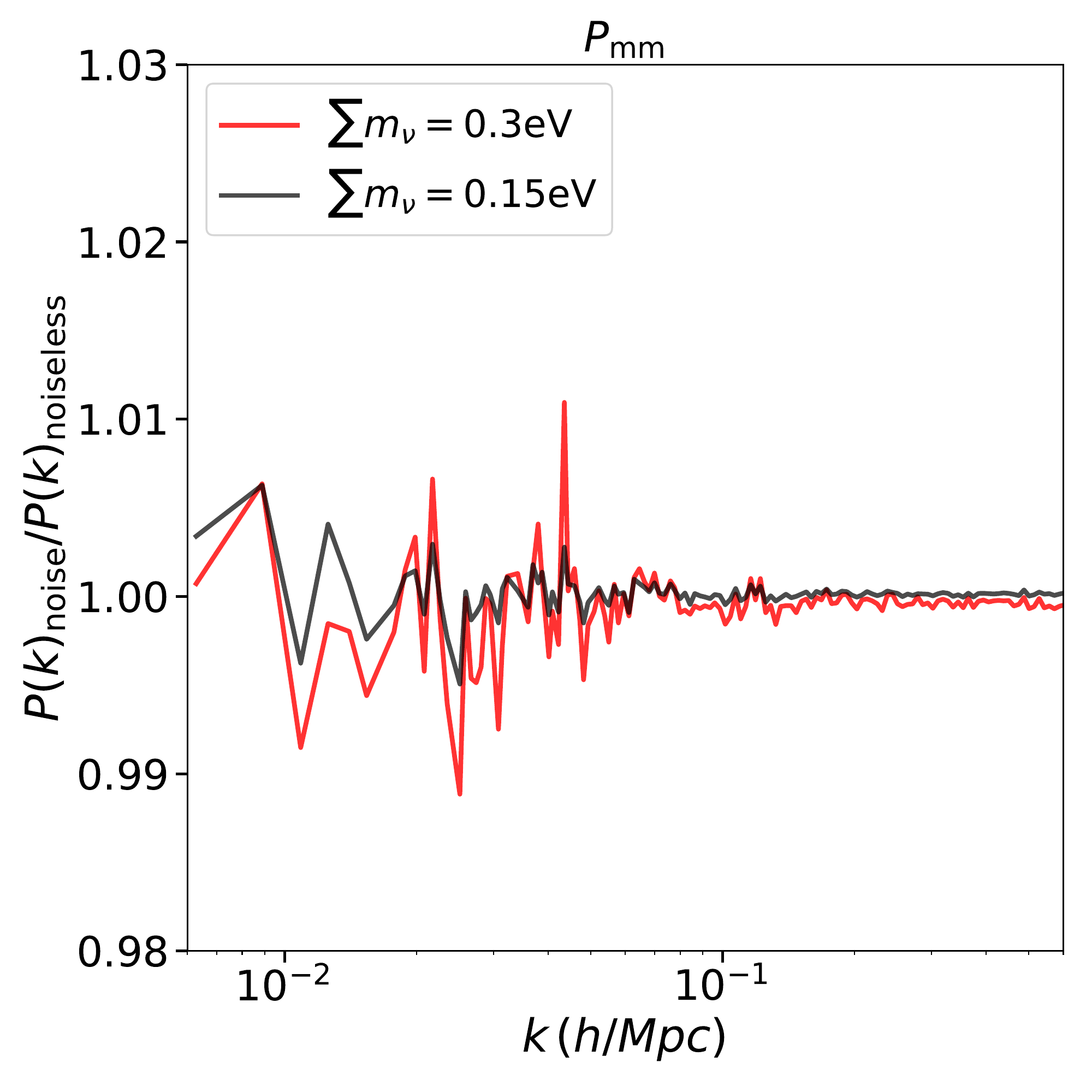}
\includegraphics[width=.45\textwidth]{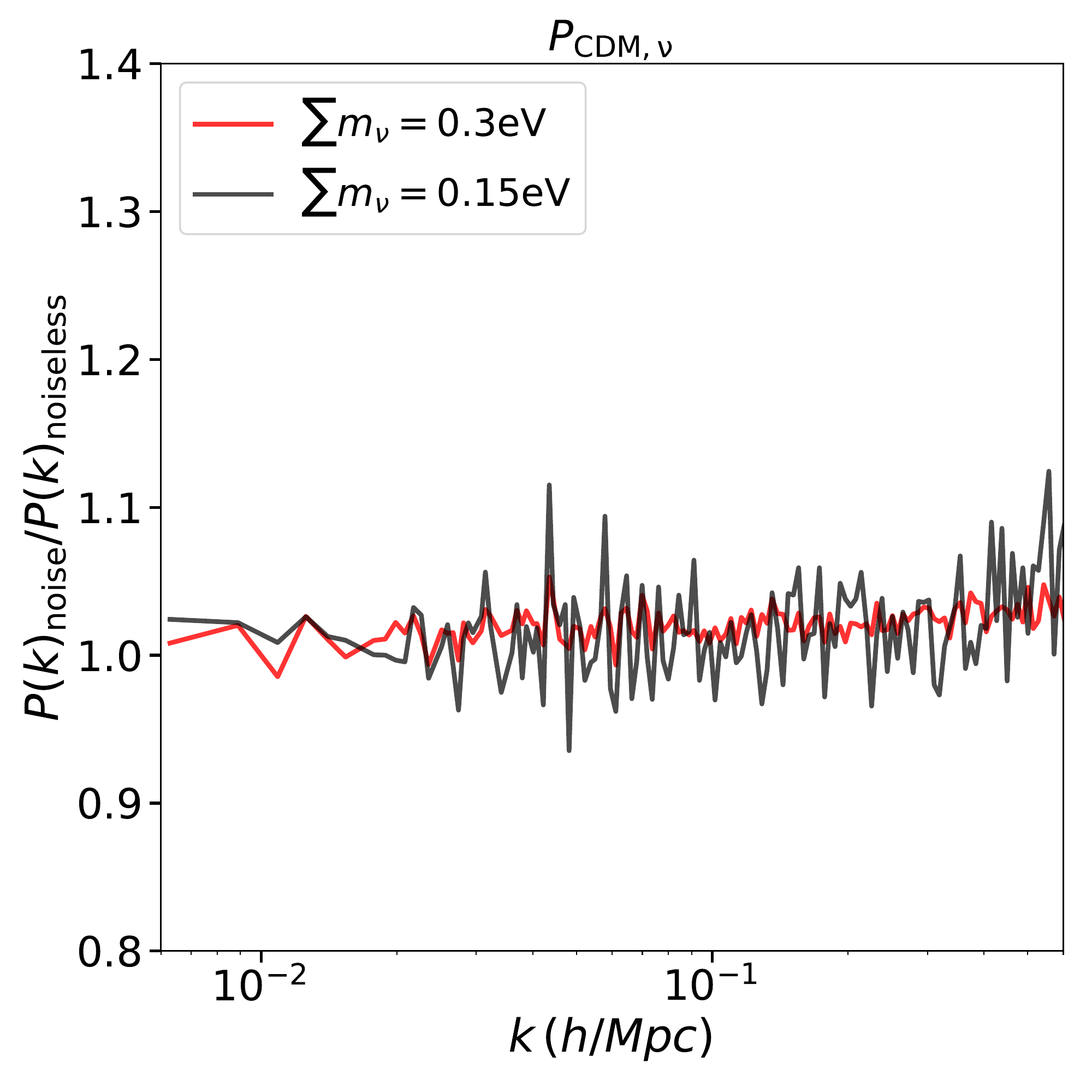}
\caption{\label{fig:a} The left panel shows the ratio of the matter power spectrum from simulations with shot noise to the same quantity from simulations without shot noise. The right hand panel shows the same for the cross-spectrum between CDM and neutrinos. The two cases considered are $\sum m_\nu = 0.15\,$eV and $\sum m_\nu = 0.3\,$eV. We find that simulations with shot noise introduce extra noise in the matter power spectrum on large scales, while the small scales are not very affected. The mean behavior of the cross correlation spectrum is also consistent between the simulations with shot noise and those without.}
\end{figure}

We first compare the ratio of the total matter power spectrum at $z=0$ from the two simulations. We find that the largest differences on the matter power spectrum from the two sets of simulations is on the larger scales, or small wavenumber. The power spectrum of the simulations with shot noise is noisy on large scales, and its magnitude is larger than that expected from cosmic variance.
%over and above that expected from sample variance. 
This is because at high redshifts the neutrino density is completely shot-noise dominated, on all scales in the box. On the other hand, in our method, the large scales do not suffer from any discreteness effects. The magnitude of the noise in the ratio of the matter power spectrum is relatively small, of the order of $1\%$, and increases with $f_\nu$, as illustrated in Fig \ref{fig:a}. We note that achieving sub-percent precision in simulations is mandatory in order to pin point percent-level effects like  the ones in \cite{Villaescusa-Navarro:2013pva, Castorina:2013wga, 2017arXiv171001310C}. Interestingly, on small scales, the matter power spectrum from the two sets of simulations agree extremely well, even though these are exactly the scales on which shot noise has the most effect on the neutrino density field. This can be understood from Eq. \ref{eq:Pkmm}. On small scales, neutrinos barely cluster so the dominant term is the CDM auto-power spectrum. The standard method induces a shot-noise term only in the neutrino-power spectrum, as the cross-power spectrum does not suffer from this. The contribution of the neutrino shot-noise to the matter power spectrum is however weighted by $f_\nu^2$, so for realistic neutrino masses this contribution to the matter power spectrum on small scales should be sub-dominant. It is however reassuring that simulations with two different setups, and very different discreteness effects in terms of sampling the neutrino distribution function agree so well for this important quantity.
%\Paco{Given the fact that} the initial conditions for the two simulations have very different discreteness effects in terms of the neutrino particles, it is reassuring that the matter power spectrum at $z=0$ match to such good precision, indicating that the results from these simulations can be trusted on small scales. 

Next we compare the cross-power spectrum between CDM and neutrinos. We find that for both mass scenarios considered, the fiducial run using the new initial conditions agree well with the cross-spectra from runs with shot noise on all scales of cosmological interest. As can be seen in the right panel of Fig. \ref{fig:a}, the mean behavior matches well, but there is some scatter between the two. Since our fiducial run was chosen such that the cross-correlation spectrum was converged with respect to various parameters of our initial conditions scheme, this implies that the random sampling of the initial distribution is able to capture the correct cross-correlation, which is dominated by the lower velocity end of the initial distribution function. It should also be pointed out that the random sampling also requires fewer particles to capture the correct cross-correlation.

Overall, we conclude that the main effect of shot noise in the neutrino density field is that the matter power spectrum becomes somewhat noisy on large scales. Using the initial condition generation scheme outlined in this paper, therefore, can help reduce the sample variance of simulations with massive neutrinos. On small scales, the effects of shot noise is hardly measurable. As long as the contribution of the neutrinos to the gravitational potential is sub-dominant to CDM, shot noise in the neutrino density field should not be correlated with fluctuations in the CDM field. 
This is exactly what we find - the cross-power spectrum of the CDM and neutrinos do not show the effects of shot noise that is clearly seen in the neutrino auto power spectrum.

\begin{figure}[tbp]
\centering 
\includegraphics[width=.45\textwidth]{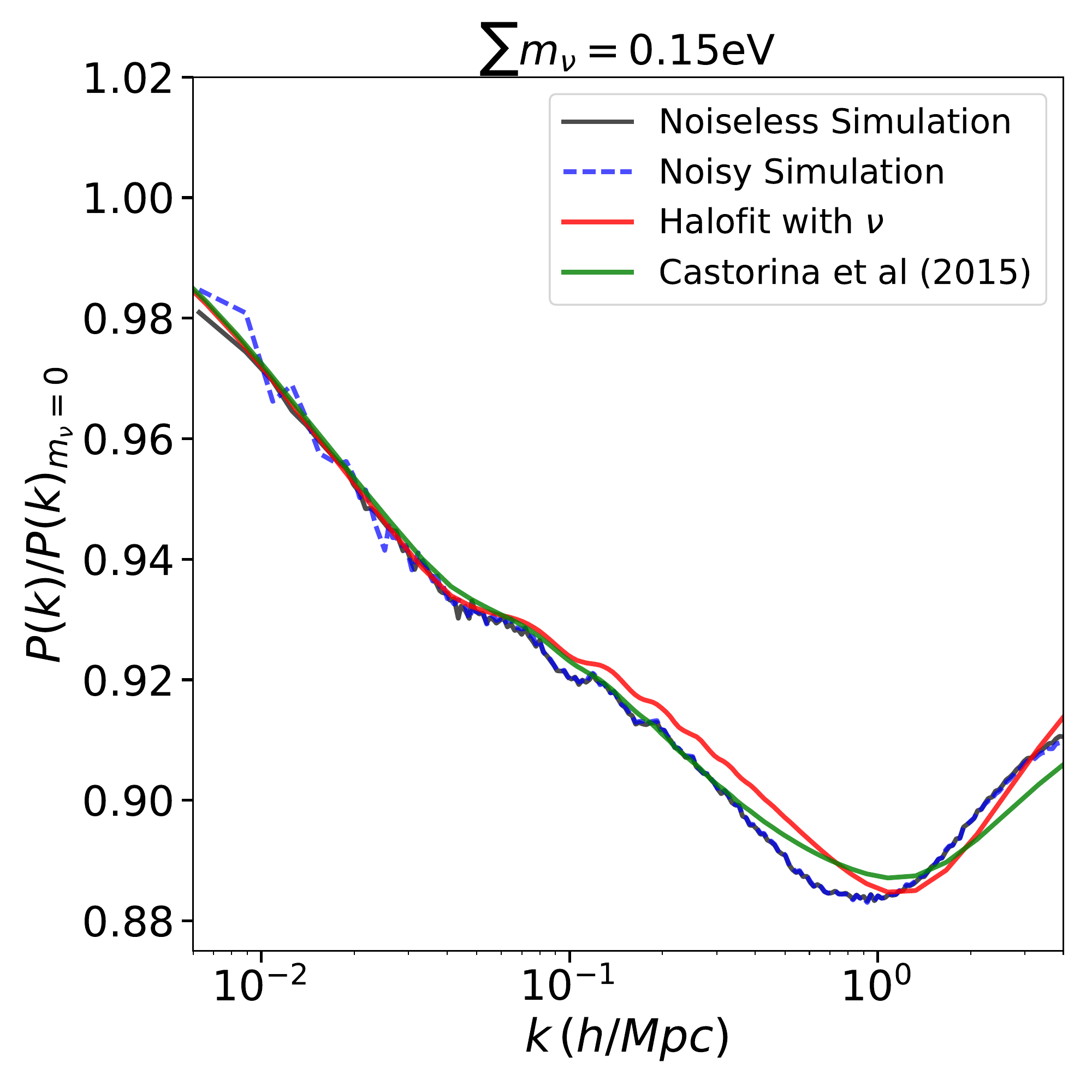}
\includegraphics[width=.45\textwidth]{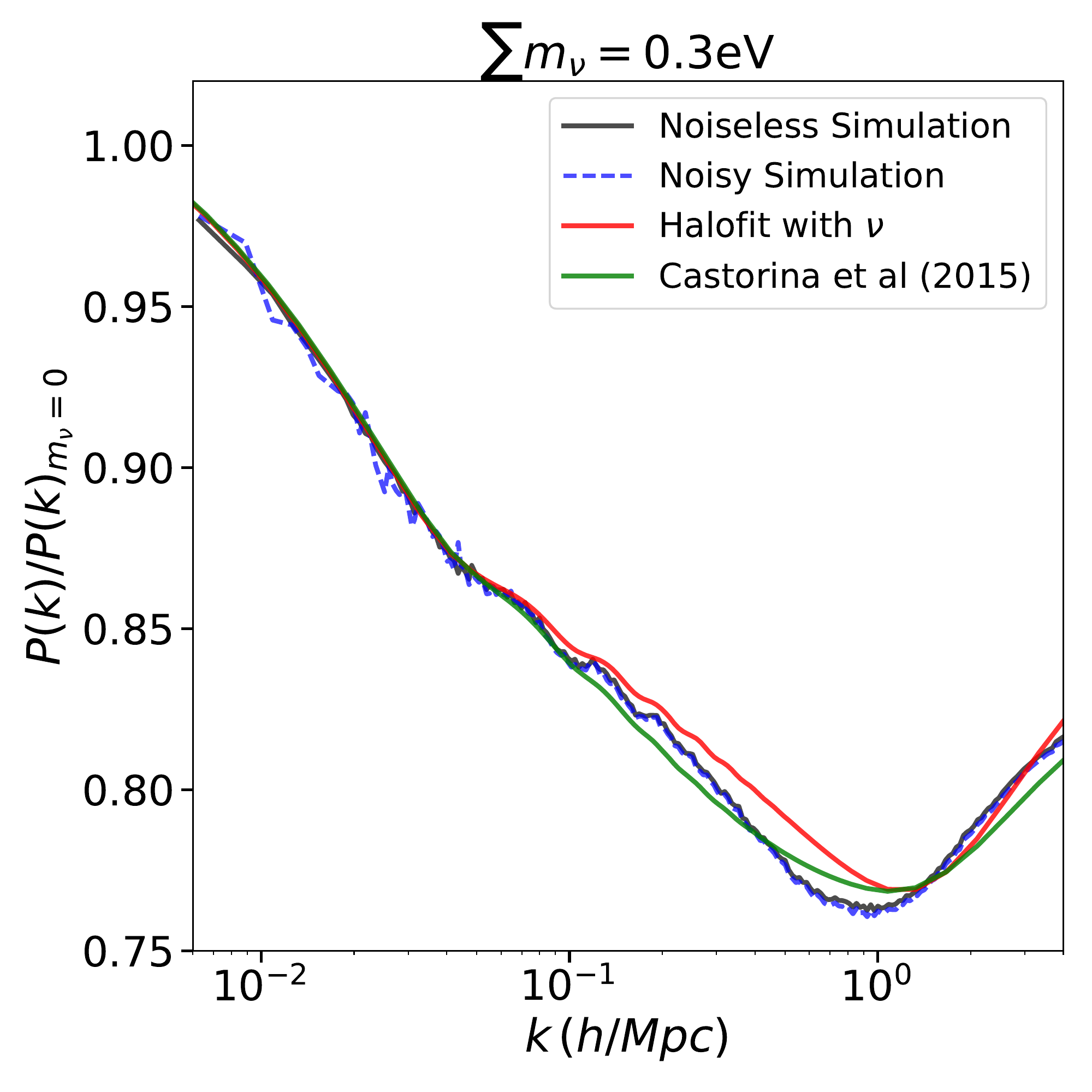}
\caption{\label{fig:d} Ratio of the fully nonlinear power spectrum at $z=0$, for cosmologies with massive neutrinos to the cosmology with no massive neutrinos. The simulation results are compared to the prediction from HALOFIT modified for massive neutrinos \cite{Bird:2011rb}, and the recipe in \cite{Castorina2015}. In the $\sum m_\nu = 0.15\,$eV case, we find very good agreement on large scales with both fitting formulas, with the latter doing somewhat better down to smaller scales. For $\sum m_\nu =0.3\,$eV, there are slightly larger differences between the results from the simulations and the two fitting formulas on smaller scales.}
\end{figure}

We now explore how the results from our simulations compare to various fitting functions and related approaches which seek to capture the effect of massive neutrinos on cosmological observables like the nonlinear matter power spectrum. The effects of massive neutrinos on the matter power spectrum was incorporated into the HALOFIT framework (\cite{Smith:2002dz,Takahashi:2012em}) in \cite{Bird:2011rb}. This method modifies the HALOFIT prescription for $\Lambda$CDM cosmologies by including terms proportional to $f_\nu$. It also takes the total matter power spectrum from linear theory as the input, rather than the cold dark matter part only. \cite{Castorina2015} suggested an alternative way to describe the effects of massive neutrinos on the total matter power spectrum. Instead of providing the linear matter power spectrum as an input to HALOFIT, the authors suggested that only the linear CDM power spectrum be used as the input to the original HALOFIT (i.e. the one without a correction for $f_\nu$). The resultant non-linear CDM power spectrum would then be combined with the linear theory predictions for the cross power spectrum of CDM and neutrinos, and the neutrino power spectrum, to yield the final matter power spectrum (see e.g. \cite{Massara_2014}). In particular,  
\begin{equation}
P_{mm}^{HF} = f_c^2 P_{cc}^{HF} + 2 f_c f_\nu P_{c\nu}^{L} + f_\nu^2 P_{\nu\nu}^{L} \,,
\end{equation}
Comparing the matter power spectrum at $z=0$ from our simulations to the fitting functions above, we find excellent agreement on large scales, seen in Fig. \ref{fig:d} for both $\sum m_\nu = 0.15$eV and $\sum m_\nu = 0.3$eV. We note that there is very little scatter on large scales, in contrast to simulations with shot noise. There are differences of a few percent on smaller scales, and we find that the prescription in \cite{Castorina2015} provides a slightly better fit out to higher values of $k$.

\begin{figure}[tbp]
\centering 
\includegraphics[width=.55\textwidth]{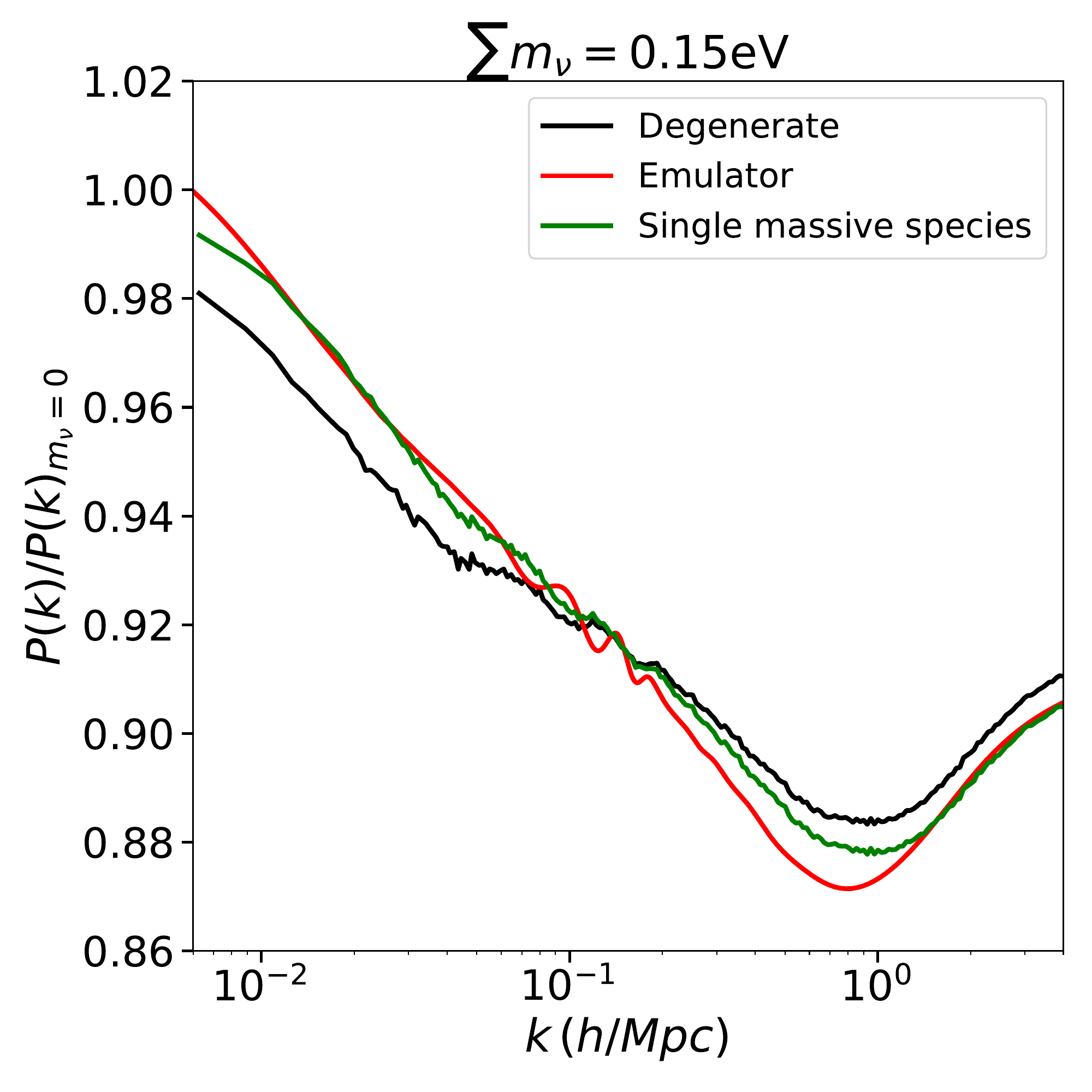}
\caption{\label{fig:e} Comparison of simulation results to the predictions of the Cosmic Emulator in \cite{Lawrence2017}. The emulator takes in only $\omega_\nu$ as an input, and assumes that this is due a single massive species, and is able to match simulations (green) with the same assumptions. However, for the same total mass, the Emulator cannot match the degenerate neutrinos scenario from the simulations (red) even on large scales.} %This suggests that in certain cases, the damping of the power spectrum is sensitive to the individual masses of neutrinos at levels which can be detected in future cosmological surveys.}
\end{figure}

We also compare the results of our simulations to the predictions from the Cosmic Emulator from \cite{Lawrence2017}. For $\sum m_\nu = 0.15$eV and for degenerate neutrino masses, we find that there is a significant difference between the two even on large scales, as shown in Fig. \ref{fig:e}. We find this difference arises from the fact that the emulator only takes $\omega_\nu$ as the input to parameterize neutrinos. $\omega_\nu$, of course, depends on the sum of the neutrino masses, $M_\nu = \sum_i m_\nu^i$, but is insensitive to the individual masses $m_i$. Our first set of simulations were run with three degenerate neutrinos, each with mass $0.05$eV. However, when we change the parameters of our simulations to run for a single massive neutrino species of mass $0.15$eV, we find good agreement between the predictions of the emulator and our simulations, as illustrated in Fig. \ref{fig:e}.

This result is interesting, since it tells us that, in principle, we could be sensitive to the individual masses of the neutrino species, along with the total mass.  This idea has been explored in literature in the context of determining the neutrino mass hierarchy \cite{DeBernardis:2009di,Jimenez:2010ev}. By comparing the different scenarios in Fig. \ref{fig:e}, we can see that the total amount of damping does depend on the sum of the masses of the neutrinos, though there are small differences in the exact amplitude. However, the scale at which the damping begins, as well as the shape of the damping on intermediate scale depends on the individual masses. This is because the mass of an individual species determines properties like the free streaming scale and transfer functions. For future surveys, where the matter power spectrum will be measured at the percent level, it will be important to explore this systematic carefully both in forecasts, as well as in the analysis. To get to the accuracy levels required for these surveys, the effect of neutrinos cannot then be captured effectively by the variation of one single parameter which only depends on the total mass of the three neutrino species, across the whole range of allowed masses and splittings. For example, if the sum of neutrino masses is close to the minimum value allowed by oscillation experiments ($\sim 0.06\,$eV), the single species is a better approximation than the degenerate mass approximation. On the other hand, if the sum of masses is much higher than the individual mass splittings, then the degenerate mass approximation is accurate, and the single species approximation will be incorrect. Since we do not want to assume, a priori, which regime we are in, the analysis needs to take into account the effects of different mass schemes for the same total mass of neutrinos.

\section{Evolution of velocity shells}
\label{sec:sheets}

We now discuss the evolution of individual neutrino shells in velocity space. A ``shell'' consists of all neutrino particles
from a given radial bin in velocity-space, or all particles with the same velocity magnitude in the initial conditions, as described in \S \ref{sec:method}.  As long as the initial particle IDs for the neutrino particles are assigned carefully, it is simple to reconstruct individual shells at any later time. By studying individual neutrino shells, we can separate out how different parts of the initial Fermi-Dirac distribution are distorted by the gravitational potential chiefly determined by the Cold Dark Matter part of the matter distribution. 

\begin{figure}[tbp]
\centering 
\includegraphics[width=1.0\textwidth]{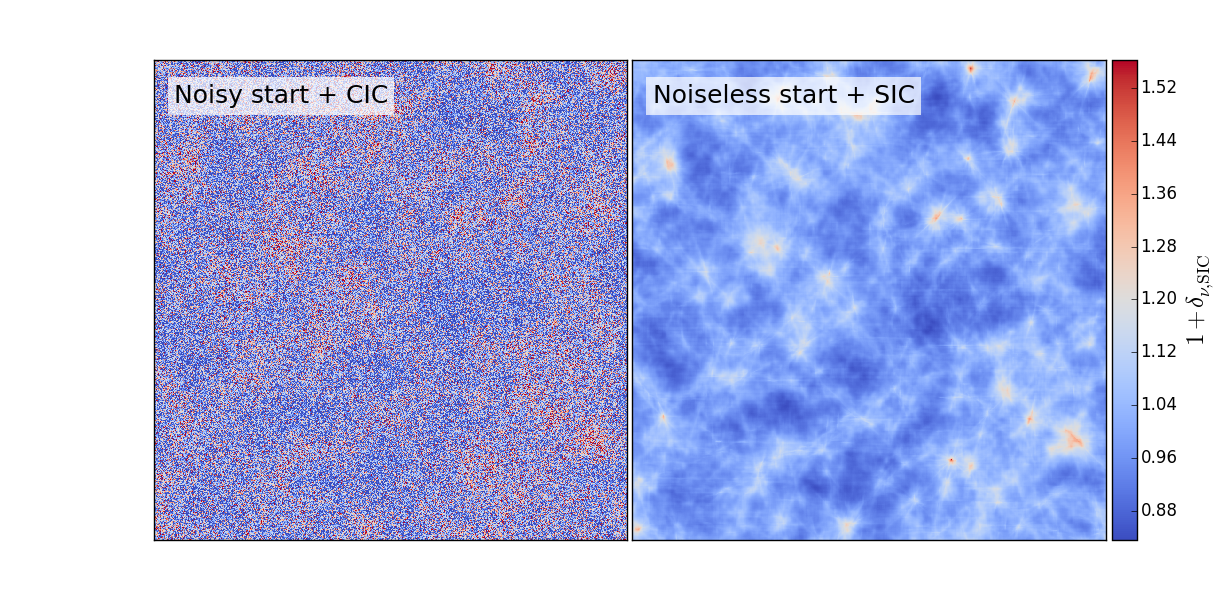}
\caption{\label{fig:picvssic} A visual comparison of the total neutrino density at $z=0$, simulated using traditional noisy initial conditions and cloud-in-cell mass deposit (CIC, run \texttt{L\_Rand\_15}, left) versus the initial conditions conditions and simplex-in-cell (SIC, run \texttt{LU\_SH10\_NS2}) techniques presented here (\S \ref{sec:sheets}, \S \ref{sec:sims}). The color bar applies only to the SIC panel; we had to arbitrarily adjust the density range in the CIC panel in order to make the overall density field visible over the noise.  The maps are projections through a $1000 \times 1000 \times 8$ Mpc/h slab, normalized to the mean density. We immediately see two major features: First is the much lower noise present in the SIC map, with physical features easily visible. Second is the pattern of ``streaks'' visible at discrete angles in the SIC map. This is a result of the finite number of bulk velocity directions available in this method for IC generation (see \S \ref{sec:method}). Neutrino streams tend to fall into massive structures preferentially along these directions. See Figure \ref{fig:timeevolution} for a map of a single one of these directional bins. We reiterate that the orderly nature of the IC generation described here allows us to uniformly discretize Lagrangian space and apply the SIC technique for interpolating mass between tracer particles.}
\end{figure}

We generate density maps for each neutrino sheet (each velocity direction within a shell, of which there are  $12\times N_{\rm side}^2\times N_{\rm shell}$ total)  using the simplex-in-cell technique (SIC). By analogy with cloud-in-cell (CIC), this is a method of mapping mass from N-body particles to a Cartesian grid.  It has been shown in \cite{Abel2011} that new physical insights can be gained by treating Cold Dark Matter as a 3-dimensional manifold in a 6-dimensional phase space, the ``phase-space sheet'' (PSS) approach. Practically speaking, this scheme demotes N-body particles to Lagrangian flow tracers and interpolates mass between the tracer particles, rather than concentrating mass at particle locations. This gives a smooth density field that is a more realistic representation of the true underlying phase-space structure.

In this application, we connect cubes of 8 N-body particles (in the initial conditions and Lagrangian space) as the mass elements in our post-processing discretization. Each neutrino sheet in the velocity discretization (as discussed in \S \ref{sec:method}) thus consists of a periodic Lagrangian grid of mass-carrying cubes. As described above, the same 8 particle IDs remain associated with the same cube for all time, so that the interpolated mass follows the bulk flow for its neutrino sheet. During the mass remapping step, each cube is refined recursively into tetrahedra using standard trilinear interpolation until each is smaller than $0.1\times\Delta x$, where $\Delta x$ is the grid resolution. Each tetrahedron is then passed to the remapping kernel, which uses the geometric intersection between a homogeneous tetrahedron and the cubical cells of a Cartesian grid to apportion mass into the final density map. This gives a smooth density field which is continuous everywhere and conserves total mass. This allows us to recover continuous density maps on a $1024^3$ grid, even after simulating only $128^3$ particles per neutrino sheet. 

The kernel to this algorithm is the geometric remapping technique of \cite{powell2015a}. 
 The remapping algorithm is designed to map mass from a tetrahedral mass element, in this case with constant mass and density inversely proportional to the volume, to a grid of cubical cells. This is done using the exactly reconstructed polyhedral intersections between each tetrahedral mass element and each cell of the grid. The resulting density field is geometrically precise, mass-conserving, and naturally anti-aliased. The algorithmic details are complex for a robust floating-point implementation; the most recent description is \cite{powell2015b}.
 Prior to this work our software package, the \texttt{Phase Sheet Intersector} (PSI), has been applied successfully in several contexts, including analysis of N-body simulations \citep{powell2015a,wojtak2016}, evolution of the Vlasov-Poisson system \citep{sousbie2015, hahn2016}, and the study of cold plasmas \citep{kh2016}.

The PSS approach as previously applied takes advantage of the coldness of the fluid in question, which can be modeled as a 3-manifold embedded in the 6D phase space. 
For neutrinos, there is the additional complication that the initial phase space does not form a single 3-dimensional manifold in 6-dimensional phase space. This arises straightforwardly from the fact that the temperatures of the neutrinos is high compared to the bulk motion. However, since neutrinos are essentially collisionless in the context of cosmology, we could decompose the initial 6-dimensional phase space into a set of 3-dimensional manifolds (the ``sheets''), which then interact with each other and to Cold Dark Matter only through gravity. As a final step, the density maps for the individual neutrino sheets are stacked into shells, with each shell containing all sheets of the same initial velocity magnitude. 

We plot the density map of neutrinos from a slice through the simulation volume on the right panel in Fig. \ref{fig:picvssic}. On the left panel, we plot the density field from the same volume, but from a simulation in which the neutrinos were initialized with random thermal velocities (\texttt{L\_Rand\_15}). Since the SIC method is not applicable to this sort of initialization, we use the traditional CIC deposition scheme to generate the density maps. The right hand panel exhibits much lower noise compared to the left, and the physical structures are clearly visible. However, since we have used a finite number of velocity directions while initializing the neutrinos, these directions remain visible in the final density map. We note that this is because we have implemented the analog of the ``horizontal streams'' discussed in \cite{kh2016}, where each velocity shell produces its own density field, and the total density field is just a sum over the density fields of all shells. Instead of using horizontal streams, one could use ``vertical streams'', also defined in \cite{kh2016}. In this case, the phase space sheets are connected along the velocity direction - i.e. all particles that started off at the same position but with different velocities now form a connected sheet. To produce the density maps of each sheet, one has to therefore interpolate in velocity space. This should lead to a smoother stacked density map despite having finite number of velocity magnitudes and directions. We plan to investigate this is detail in future work.

\begin{figure}[tbp]
\centering 
\includegraphics[width=1.0\textwidth]{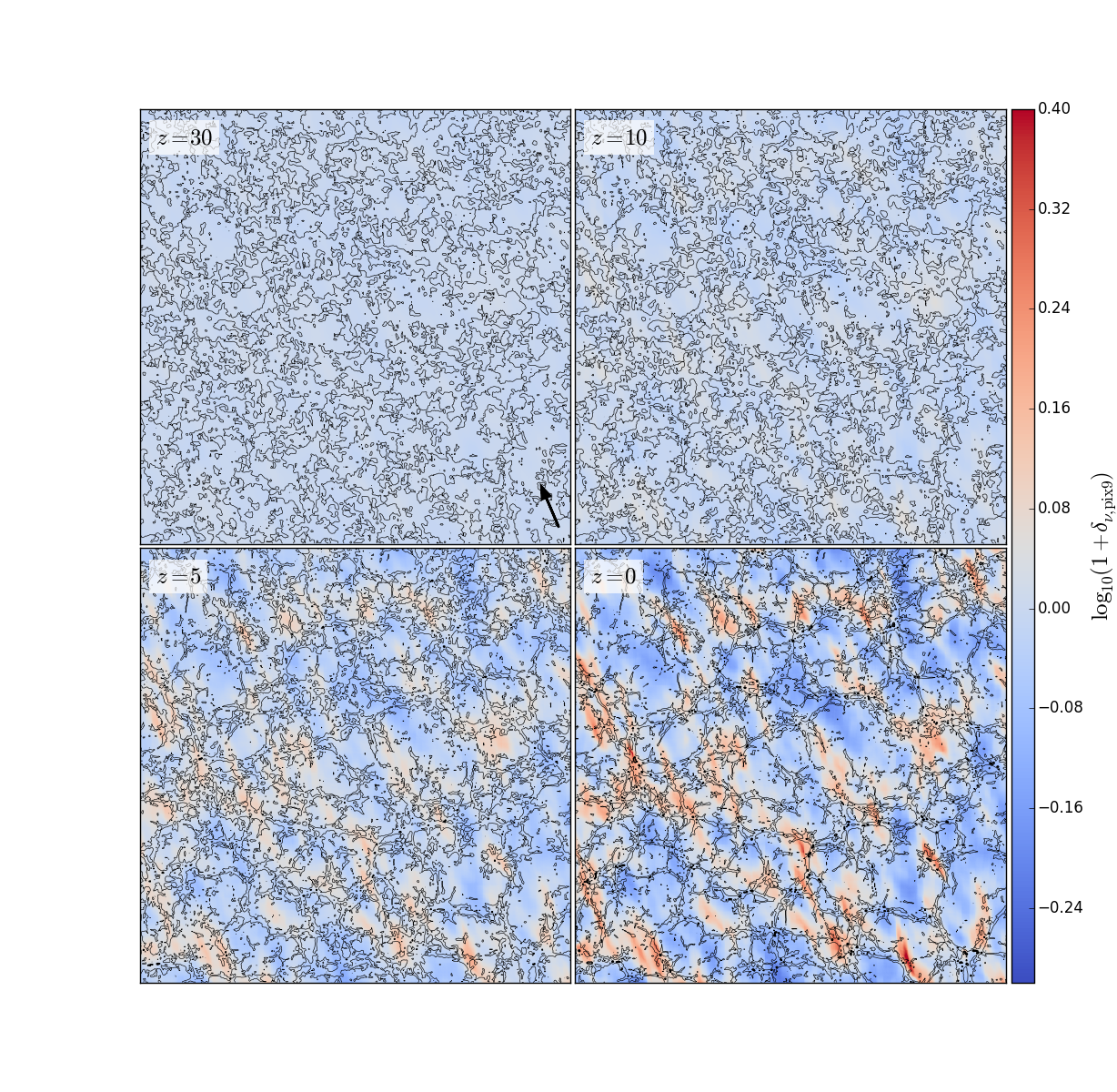}
\caption{\label{fig:timeevolution} Time-evolution of the neutrino density for a single initial velocity direction; this is the sum of all neutrino shells (velocity magnitude bins) for \texttt{HEALPIX} pixel 9 from run \texttt{LU\_SH10\_NS2} (see \S \ref{sec:method}, \S \ref{sec:sims}). The bulk velocity is shown by the arrow in the upper-left frame.  The CDM density is given by the black contours. Thin, thick contours correspond to $\log_{10}(1+\delta_\mathrm{CDM})=0,1$ respectively.
All maps are projections through a $250 \times 250 \times 8$ Mpc/h slab, normalized to the mean density. We use the simplex-in-cell (SIC) method (\S \ref{sec:sheets}) to generate smooth and accurate density maps from Gadget particle data.}
\end{figure}

In Fig. \ref{fig:timeevolution}, we present the time evolution of density field, but this time using only one of the velocity directions chosen using \texttt{HEALPIX}. To generate this density field, we have summed over the different velocity magnitudes. The velocity direction is shown with the black arrow in the upper left frame. As expected, the clustering in the neutrinos increases with time, as is correlated with the clustering of the CDM component shown in black. Since the neutrinos fall in along one direction, the visible structures at $z=0$ are all elongated along this direction.

%It is by examining these shells that we recover the dependence of the $z=0$ power spectrum for each shell on its initial velocity.

\begin{figure}[tbp]
\centering 
\includegraphics[width=.45\textwidth]{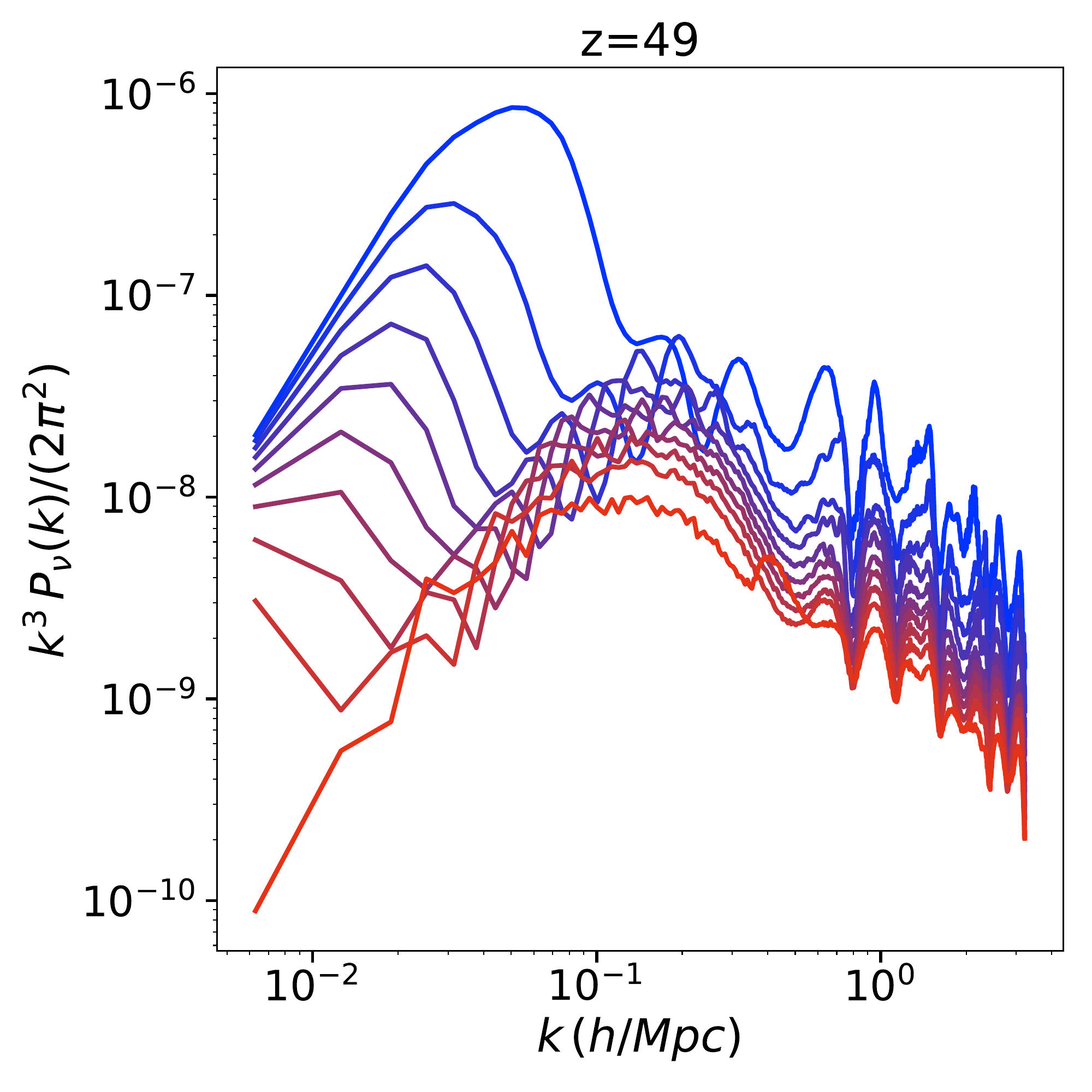}
\includegraphics[width=.45\textwidth]{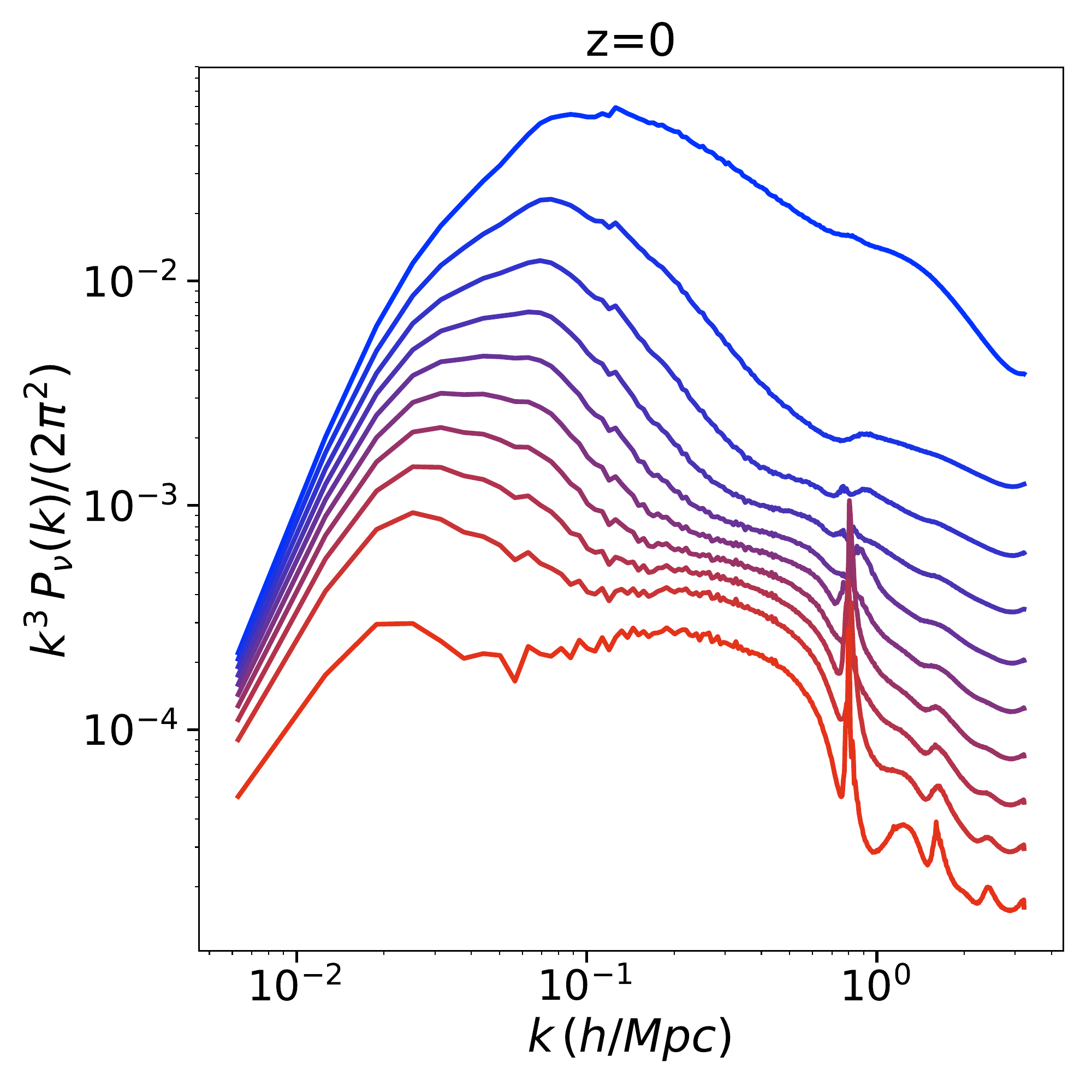}
\caption{\label{fig:b} Dimensionless power of different shells of neutrinos at $z=49$ (left panel) and at $z=0$ (right panel) from the simulation \texttt{LU\_SH10\_NS2}. The slower neutrino shells always have higher power, and their power at small scales (large $k$) is enhanced at late times. The feature in the faster shells on the right panel near $k = 0.8\,$h/Mpc is an artifact of the original grid size. For the slower shells, the physical power is larger than the grid artifact, and hence does not appear for them. Blue corresponds to neutrino shells with slow bulk velocity, and red corresponds to fast-moving shells.}
\end{figure}

Using the density maps, we study the evolution of the dimensionless power ($k^3\,P(k)/(2\pi^3)$) for each shell at different redshifts - we concentrate on $z=49$ and $z=0$ as being representative of high redshift and low redshift behavior, respectively. Even though all the sheets started off with the same initial conditions at $z=99$, the left panel of Fig. \ref{fig:b} shows that, even at high redshifts, the power in each shell is different. As expected, the power in the slower shells is higher at a given redshift when compared to the power in faster shells. This is a direct consequence of the fact that the shells have different thermal velocities, and therefore, a different effective ``sound speed''.  However, at such high redshifts, the sound speed of each shell is given roughly by the unperturbed Fermi-Dirac distribution. The sound speed of the shells therefore redshift in the same manner, and the offsets between the different sheets seen in Fig. \ref{fig:b} (left panel) do not grow significantly at high redshifts. This fits well with our finding in \S \ref{sec:conv} that we do not need many neutrino shells at early times to follow the evolution correctly. 

At low redshifts, however, we expect more complicated behavior in terms of the evolution of the neutrino shells. The largest scales still evolve in the same way for the different shells. But on small scales, the slower shells are expected to respond more strongly to the gravitational potential which is mainly sourced by CDM, whereas the faster shells are still relatively unaffected. We see this in the right panel of Fig. \ref{fig:b}, where the shape of the power spectrum of the slowest shell is quite different on scales of $k \sim 0.1\,$h/Mpc to $k\sim 1\,$h/Mpc compared to the faster shells. 
%We apply this technique to study the evolution of the different shells at high redshift, $z=49$, from the simulation \texttt{E\_SH5\_NS4}. We do this in terms of the power spectrum of each individual shell. We find that, as expected, the power spectra of the shells are different from each other, even though the initial conditions for all the shells were the same except for their thermal velocities. This difference of thermal velocities leads to a difference in the ``sound speed'' of each individual shell. However, at such high redshifts, the sound speed of each shell is given roughly by the unperturbed Fermi-Dirac distribution. The sound speed of the shells therefore redshift in the same manner, and the offsets between the different sheets seen in Fig. \ref{fig:b} do not grow significantly at high redshifts. This fits well with our finding in \S \ref{sec:conv} that we do not need many neutrino shells at early times to follow the evolution correctly.

%A low redshifts, however, we expect more complicated behavior in terms of the evolution of the neutrino shells. The slower shells are expected to respond more strongly to the gravitational potential which is mainly sourced by CDM, whereas the faster shells are still relatively unaffected. 

\begin{figure}[tbp]
\centering 
\includegraphics[width=.8\textwidth]{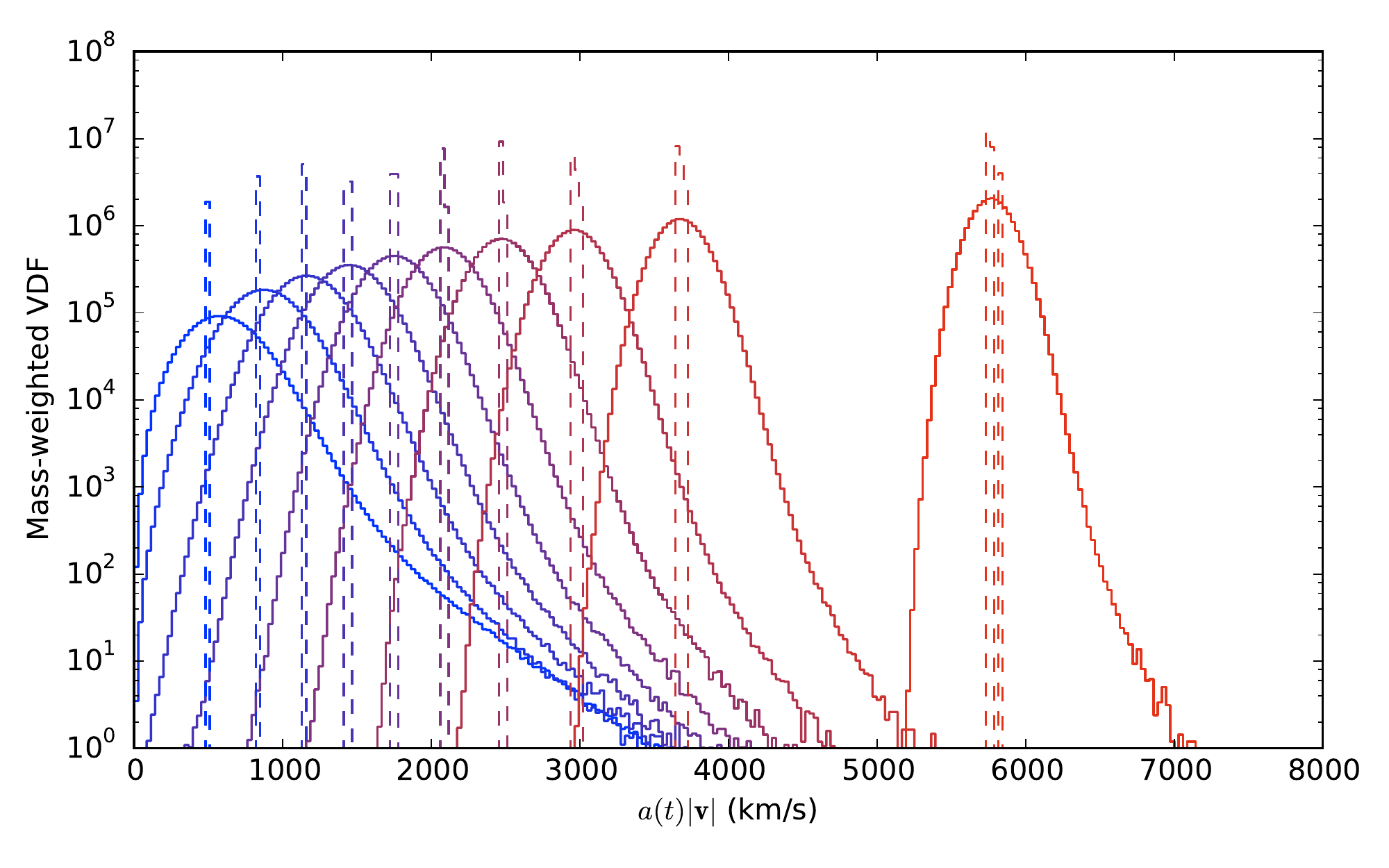}
\caption{\label{fig:vdf} The velocity distribution function of each neutrino shell, at $z=0$ (solid) and the initial conditions (dashed). The initial velocity-space distribution reflects spherical ``shells'' in velocity-space that each sample a different energy range of the initial Fermi-Dirac distribution (see Figure \ref{fig:l}). Blue corresponds to neutrino shells with slow bulk velocity, and red corresponds to fast-moving shells. }
\end{figure}

The effect of the gravitational potential on different shells at low redshift becomes clear when we examine the velocity distribution function (VDF) of the shells from the simulation \texttt{LU\_SH10\_NS2}. When the initial conditions are generated, the distribution function of each shell is a Dirac delta function at the velocities given by Eq. \ref{eq:shell_vel}, as can be seen by the dashed curves in Fig. \ref{fig:vdf}. At $z=0$, the distribution functions are modified, as illustrated by the solid curves. In particular, the distribution functions of all the shells broaden, with the slower shells broadening much more than the faster ones, in agreement with previous works \citep{2013JCAP...03..019V}. It is interesting to note that for the fastest shell, the total amount of broadening about the initial velocity is roughly $10^3\,$km/s, roughly the velocity dispersion of the largest galaxy clusters. The slow shells have a significant fraction of particles which get sped up by many multiples of the initial velocity with which they started.

\begin{figure}[tbp]
\centering 
\includegraphics[width=.8\textwidth]{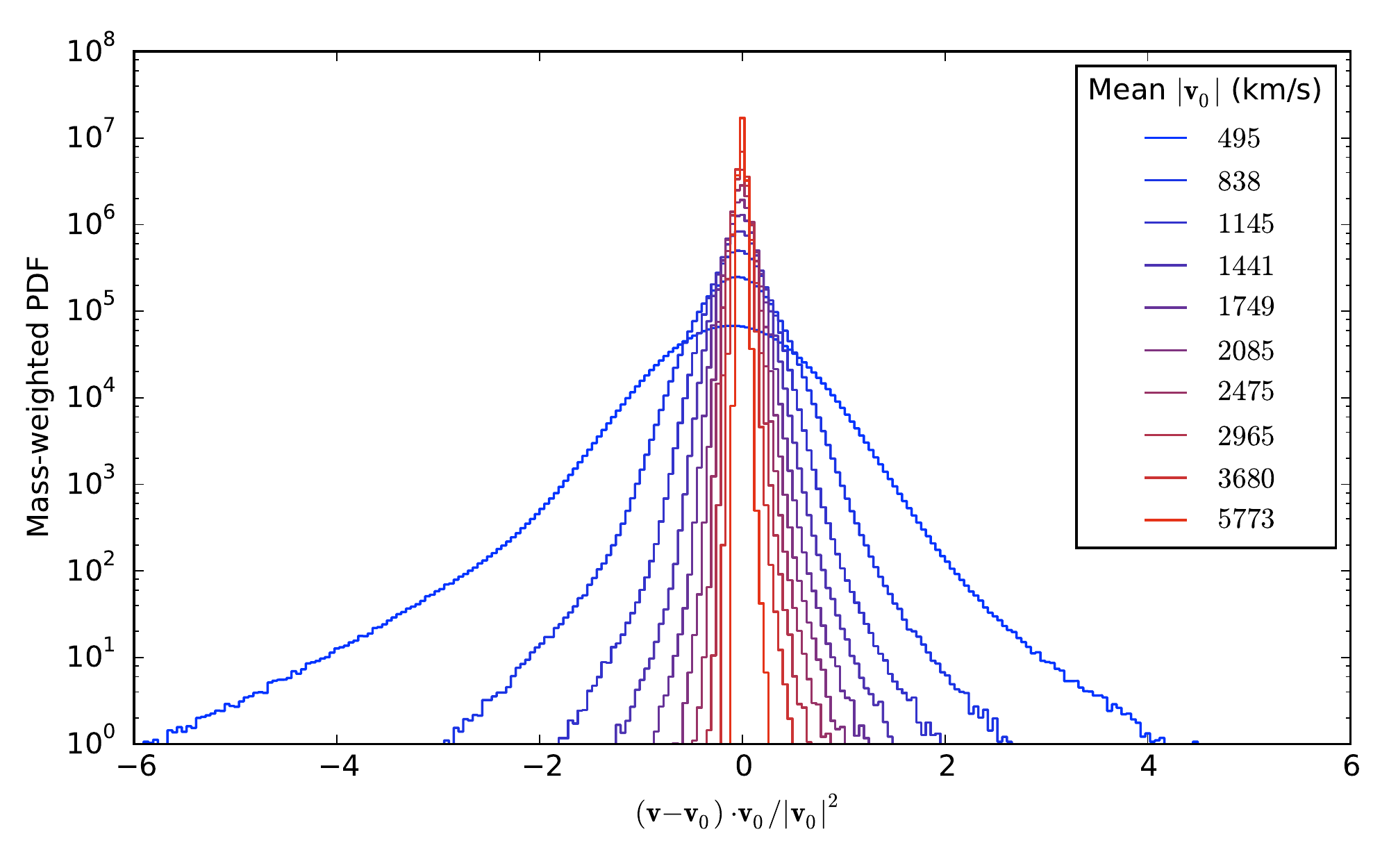}
\caption{\label{fig:deltavdf} The distribution of changes in velocity for each neutrino shell. Note that the low-velocity shells are much more susceptible to broadening of the VDF due to interactions with massive galaxy clusters. }
\end{figure}

To understand these high velocity tails, we compute the quantity $\left(\mathbf{v} - \mathbf{v_0}\right).\mathbf{v_0}/\left|\mathbf{v_0}\right|^2$ for particles in every shell, and plot the mass-weighted distribution of this quantity in Fig. \ref{fig:deltavdf}. Here, $\mathbf{v_0}$ represents the original velocity of the neutrino particle, but whose magnitude has been scaled to account for the decay of peculiar velocities with redshift. For neutrino particles which are not captured by the captured by gravitational potentials, but are instead just sped up or slowed down by a small amount along the original direction of motion of the neutrino particles, we expect $\left(\mathbf{v} - \mathbf{v_0}\right).\mathbf{v_0}/\left|\mathbf{v_0}\right|^2$ to be relatively small. This is clearly illustrated by the fastest neutrino shell, which shows the smallest deviation from $0$ in Fig. \ref{fig:deltavdf}. For particles which are actually captured by the gravitational potentials, there are two effects. The first is that the magnitude of the total velocity of the particles can change by an amount much larger than the magnitude of the original velocity of the particle. The second effect is that the direction of the velocity of the particle at $z=0$ is not correlated with the original velocity direction of the particle when the initial conditions were generated. The first effect is illustrated by the large broadening of the distribution for the slower neutrino shells. The second effect shows up as the asymmetry of the distribution function about $0$ - i.e. particles which have been captured have, at some point in their trajectory, changed their direction of motion with respect to the original velocity.

\begin{figure}[tbp]
\centering 
\includegraphics[width=.8\textwidth]{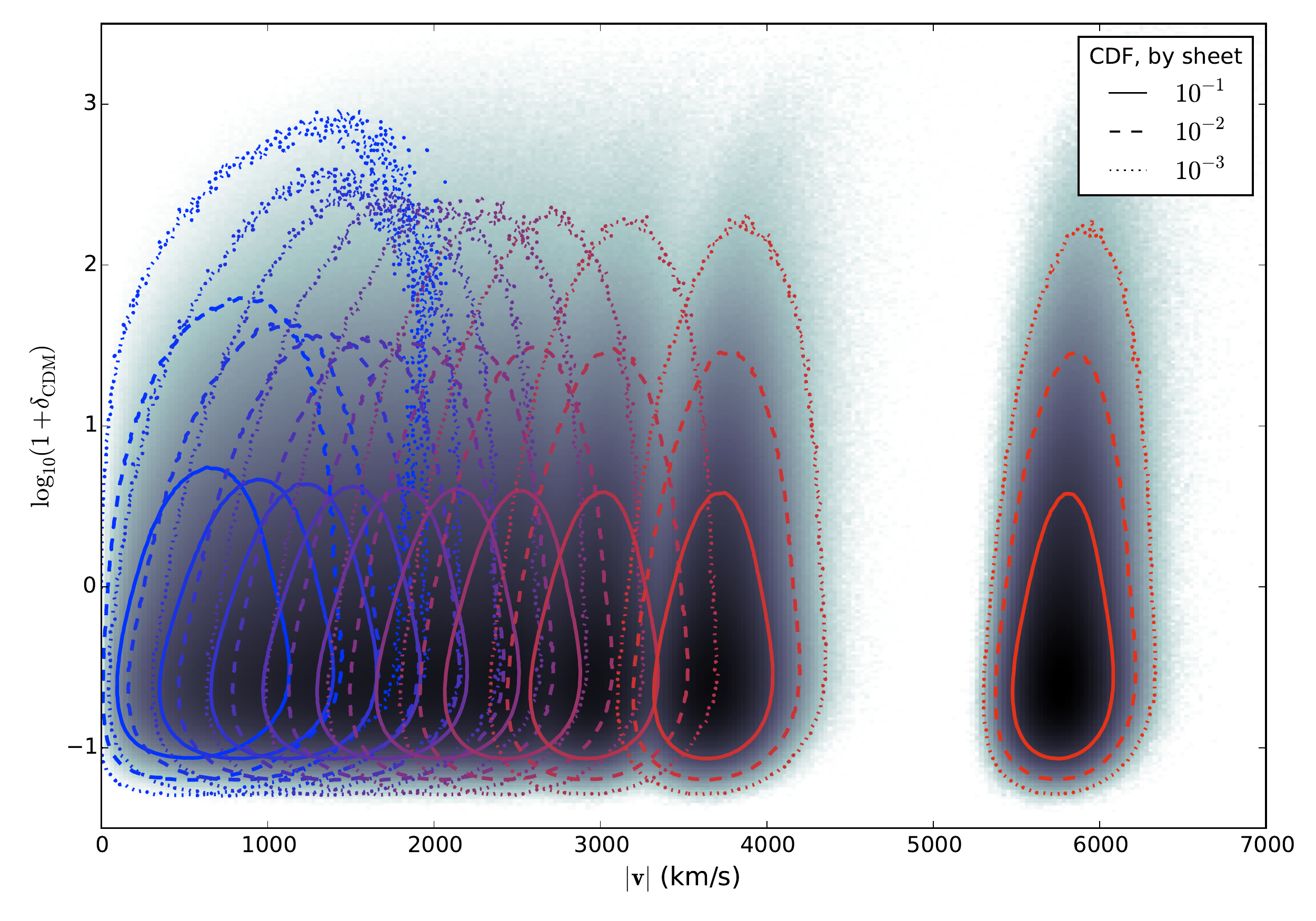}
\caption{\label{fig:vdf_vs_rho} The mass-weighted distribution of neutrinos in the space of neutrino velocities and CDM density at $z=0$. Contours are drawn for the total mass fraction in each sheet \emph{outside} of each contour. The grayscale image is logarithmic in the PDF, and is meant to guide the eye through the tails of the total distribution. Note that the low-velocity shells tend to spread towards regions with higher CDM density.}
\end{figure}

To check that the large change in the velocities of neutrino particles occurs in regions where the CDM density, and therefore the potential, is much larger than the background value, we plot the local CDM density of each neutrino particle vs. its own velocity in Fig. \ref{fig:vdf_vs_rho}. Once again, for the high velocity shells, the local CDM density does not produce a significant change in the equal mass contours plotted in Fig. \ref{fig:vdf_vs_rho}. For the slower shells on the other hand, we find that the shapes of the equal mass contours are very different in underdense vs. overdense regions of CDM. For a given shell, many more of its fastest particles can be found in regions of the simulation volume where the CDM density is high compared to those regions of low density.

This technique therefore enables us to follow the evolution of the distribution function of the neutrinos in different environments in a systematic manner. Since the original thermal velocities for all neutrino particles in the simulation come from a finite number of pre-determined values, it is easy to follow how different parts of the original Fermi-Dirac distribution are modified as result of gravitational interactions with CDM. It is interesting to note that even for neutrino masses as low as $0.05\,$eV for individual species, there is significant clustering of the slower shells in our simulations. We found this in the cross-correlation power spectrum discussed in \S \ref{sec:results}, as well as in the broadening of the distribution function shown in Figs. \ref{fig:vdf}, \ref{fig:deltavdf}, and \ref{fig:vdf_vs_rho}. Studying this clustering of neutrinos can be useful in the context of experiments like PTOLEMY \cite{Betts:2013uya}, which attempt to make direct detections of the cosmic neutrino background.

\section{Summary and Discussion}
\label{sec:disc}
We have presented a new method for generating initial conditions for   cosmological simulations with massive neutrinos as an extra set of N-body particles. Previous methods for generating such initial conditions were plagued by Poissonian shot noise arising from the streaming of neutrino particles. The method presented in this paper is able to eliminate this shot noise by choosing appropriate initial conditions. We achieve this by sampling the Fermi-Dirac distribution, from which the thermal velocities of the neutrinos are drawn, in a regular manner, as opposed to randomly sampling it, as was done previously. We divide the Fermi-Dirac into intervals to choose the magnitude of the velocity of neutrinos, and the directions of the neutrino particles are chosen following the \texttt{HEALPIX} algorithm. The crucial point here is that the same values are repeated over all points on the grid on which the initial conditions are generated. This ensures that the number of neutrino particles moving in and out of various regions of the simulation volume is no longer random - this being the source of the Poisson noise seen in previous simulations. 

Of course, this method has its own discreteness effects - namely the fact that we choose finite number of points to sample the initial distribution function and a finite number of directions in which we add the thermal velocities to the neutrino particles. We have investigated how the different choices of parameters in our current method affect the high-redshift neutrino power spectrum, and the matter power spectrum and CDM-neutrino cross correlation spectrum at $z=0$. At high redshifts, as long as the initial distribution is roughly isotropic, the simulations can match linear theory predictions for the neutrino power spectrum. The number of shells into which we divide the Fermi-Dirac distribution is not very important in this regime. At low redshifts, while considering the matter power spectrum, the results from the simulations are well-converged with respect to various parameters of our method. However, when we consider the cross spectrum, which measures how correlated the clustering of the neutrinos is with respect to CDM, we find that we need to sample the lower end of the initial Fermi-Dirac distribution well in order to achieve convergence. We find that the slow-moving neutrinos contribute most to the cross correlation, disproportionate to the initial phase space volume that they occupy. 

When we compare the results of the simulations run with these initial conditions to simulations with shot noise in the neutrinos, we can essentially isolate the effects of shot noise on various observables. As mentioned earlier, this provides an extremely valuable check on the results of various neutrino simulations run using the N-body method. We note that while the new method does use more neutrino particles than is used currently, the new method is far more effective at removing shot noise than by just increasing the number of particles while using existing methods to assign thermal velocities to neutrino particles. For example, using $\sim 1000^3$ neutrino particles, this method is able to better the shot noise levels of the largest current neutrino simulations \cite{Yu:2016yfe} ($\sim 13000^3$ neutrino particles) by at least three orders of magnitude. For the matter power spectrum, we find that there is virtually no difference between the two sets of simulations on small scales, even though this is exactly where the shot noise issue is most dominant for neutrinos. This confirms that the neutrino power spectrum is so damped on these scales that even large amounts of noise in it will be sub-dominant, and not show up in observables like the matter power spectrum. However, we find that using the ``noiseless'' initial conditions reduces the noise on large scales. This is because in the ``noisy'' simulations, even the largest scales in the simulation box are affected by shot noise at early times, and the effect of this is imprinted onto the late time matter power spectrum. Looking ahead, this reduction in the noise will useful in beating down the sample variance of massive neutrino simulations. In terms of the cross-correlation, we find that the random sampling of the initial Fermi-Dirac distribution is quite effective in being able to follow the clustering of the low end of the velocity distribution.

Given that our results mostly match those from previous simulations, it is not surprising that we also find a close match between our results and various fitting formulas and emulators in literature trying to capture the effects of massive neutrinos on the nonlinear matter power spectrum \cite{Bird:2011rb,Castorina2015,Lawrence2017}. This is especially true at large scales. We also point out that the damping of the power spectrum can be sensitive to what the individual masses of the neutrino mass eigenstates are, in addition to the total mass. Individual neutrino masses set the free streaming mass, which determines the scale at which the damping due to neutrinos starts becoming effective. While this has been known, we point out that in certain situations, the difference between a single massive neutrino species, and the degenerate scenario, for the same total mass, can produce differences which could be measurable in future surveys \cite{DESI,EUCLID,WFIRST,LSST}. This means that using a single parameter, such as the total mass, to capture the effects of neutrino mass in the forecasts and analysis for these surveys may no longer be sufficient.

By suitably ordering the particle IDs in the initial conditions, we can also follow the evolution of the individual ``shells'' into which we divide the Fermi-Dirac distribution. This is done by modifying the ``phase-space sheet'' outlined in \cite{Abel2011} to account for the initial velocity distribution function of neutrinos. We find that even for light neutrino species, there can be significant clustering for the slow shells. This is especially true in regions where the density of CDM itself is large - i.e. inside dark matter halos. This method, therefore, can be extremely useful in understanding the modification of the distribution function of neutrinos as a function of the local environment. It is also useful to study the enhancement of the local neutrino density in dense environments, in the context of experiments which seek to detect the cosmic neutrino background \cite{Betts:2013uya}. In this context, we point out that in this paper, we have used a scheme analogous to the ``horizontal streams'' outlined in \cite{kh2016}. While this scheme already outperforms the existing methods for the generation of initial conditions for neutrino particles, some of the discreteness effects associated with it (such as the discrete initial velocity magnitudes and directions) can be alleviated by implementing ``vertical streams'' scheme outlined in \cite{kh2016}. In that scheme, density reconstructions involve interpolations in velocity space, smoothing out the effects of having a finite number of sample points. We plan to explore these directions in more detail in future studies.

\acknowledgments

We would like to thank Matteo Viel and Volker Springel for permission to use the cosmological code Gadget-3. All simulations for this paper were run on the Sherlock cluster at Stanford University. We used \texttt{GenPK} \cite{2017ascl.soft06006B} extensively to generate different power spectra in this paper. We also used \texttt{pyGadgetReader} \cite{2014ascl.soft11001T} in our analysis pipeline to read different snapshots from simulations. This work was performed in part under DOE Contract DE-AC02-76SF00515 and benefitted from the Stanford Research Computing Center.


\bibliography{ics}

\providecommand{\href}[2]{#2}\begingroup\raggedright\begin{thebibliography}{10}

\bibitem{SuperK98}
{\scshape Super-Kamiokande} collaboration, Y.~Fukuda et~al., \emph{{Evidence
  for oscillation of atmospheric neutrinos}},
  \href{http://dx.doi.org/10.1103/PhysRevLett.81.1562}{\emph{Phys. Rev. Lett.}
  {\bf 81} (1998) 1562--1567}, [\href{http://arxiv.org/abs/hep-ex/9807003}{{\tt
  hep-ex/9807003}}].

\bibitem{SNO2001}
{\scshape SNO Collaboration} collaboration, Q.~R. Ahmad, R.~C. Allen, T.~C.
  Andersen, J.~D. Anglin, G.~B\"uhler, J.~C. Barton et~al., \emph{Measurement
  of the rate of
  ${\ensuremath{\nu}}_{e}+\mathit{d}\ensuremath{\rightarrow}\mathit{p}+\mathit{p}+{\mathit{e}}^{-}$
  interactions produced by $^{8}b$ solar neutrinos at the sudbury neutrino
  observatory},
  \href{http://dx.doi.org/10.1103/PhysRevLett.87.071301}{\emph{Phys. Rev.
  Lett.} {\bf 87} (Jul, 2001) 071301}.

\bibitem{K2K2003}
{\scshape K2K Collaboration} collaboration, M.~H. Ahn, E.~Aliu, S.~Andringa,
  S.~Aoki, Y.~Aoyama, J.~Argyriades et~al., \emph{Measurement of neutrino
  oscillation by the k2k experiment},
  \href{http://dx.doi.org/10.1103/PhysRevD.74.072003}{\emph{Phys. Rev. D} {\bf
  74} (Oct, 2006) 072003}.

\bibitem{DayaBay}
F.~P. An, J.~Z. Bai, A.~B. Balantekin, H.~R. Band, D.~Beavis, W.~Beriguete
  et~al., \emph{Observation of electron-antineutrino disappearance at daya
  bay}, \href{http://dx.doi.org/10.1103/PhysRevLett.108.171803}{\emph{Phys.
  Rev. Lett.} {\bf 108} (Apr, 2012) 171803}.

\bibitem{Dodelson2003book}
S.~{Dodelson}, \emph{{Modern cosmology}}.
\newblock {Academic Press}, 2003.

\bibitem{Hu1997}
W.~Hu, D.~J. Eisenstein and M.~Tegmark, \emph{{Weighing neutrinos with galaxy
  surveys}}, \href{http://dx.doi.org/10.1103/PhysRevLett.80.5255}{\emph{Phys.
  Rev. Lett.} {\bf 80} (1998) 5255--5258},
  [\href{http://arxiv.org/abs/astro-ph/9712057}{{\tt astro-ph/9712057}}].

\bibitem{Lesgourgues2006}
J.~Lesgourgues and S.~Pastor, \emph{{Massive neutrinos and cosmology}},
  \href{http://dx.doi.org/10.1016/j.physrep.2006.04.001}{\emph{Phys. Rept.}
  {\bf 429} (2006) 307--379},
  [\href{http://arxiv.org/abs/astro-ph/0603494}{{\tt astro-ph/0603494}}].

\bibitem{Planck2013}
{\scshape Planck} collaboration, P.~A.~R. Ade et~al., \emph{{Planck 2013
  results. XVI. Cosmological parameters}},
  \href{http://dx.doi.org/10.1051/0004-6361/201321591}{\emph{Astron.
  Astrophys.} {\bf 571} (2014) A16},
  [\href{http://arxiv.org/abs/1303.5076}{{\tt 1303.5076}}].

\bibitem{Planck2015}
{\scshape Planck} collaboration, P.~A.~R. Ade et~al., \emph{{Planck 2015
  results. XIII. Cosmological parameters}},
  \href{http://dx.doi.org/10.1051/0004-6361/201525830}{\emph{Astron.
  Astrophys.} {\bf 594} (2016) A13},
  [\href{http://arxiv.org/abs/1502.01589}{{\tt 1502.01589}}].

\bibitem{Sherwin2016}
B.~D. Sherwin et~al., \emph{{Two-season Atacama Cosmology Telescope polarimeter
  lensing power spectrum}},
  \href{http://dx.doi.org/10.1103/PhysRevD.95.123529}{\emph{Phys. Rev.} {\bf
  D95} (2017) 123529}, [\href{http://arxiv.org/abs/1611.09753}{{\tt
  1611.09753}}].

\bibitem{deHaan2016}
{\scshape SPT} collaboration, T.~de~Haan et~al., \emph{{Cosmological
  Constraints from Galaxy Clusters in the 2500 square-degree SPT-SZ Survey}},
  \href{http://dx.doi.org/10.3847/0004-637X/832/1/95}{\emph{Astrophys. J.} {\bf
  832} (2016) 95}, [\href{http://arxiv.org/abs/1603.06522}{{\tt 1603.06522}}].

\bibitem{Palanque-Delabrouille:2015pga}
N.~Palanque-Delabrouille et~al., \emph{{Neutrino masses and cosmology with
  Lyman-alpha forest power spectrum}},
  \href{http://dx.doi.org/10.1088/1475-7516/2015/11/011}{\emph{JCAP} {\bf 1511}
  (2015) 011}, [\href{http://arxiv.org/abs/1506.05976}{{\tt 1506.05976}}].

\bibitem{Palanque-Delabrouille:2014jca}
N.~Palanque-Delabrouille et~al., \emph{{Constraint on neutrino masses from
  SDSS-III/BOSS Ly$\alpha$ forest and other cosmological probes}},
  \href{http://dx.doi.org/10.1088/1475-7516/2015/02/045}{\emph{JCAP} {\bf 1502}
  (2015) 045}, [\href{http://arxiv.org/abs/1410.7244}{{\tt 1410.7244}}].

\bibitem{Abbott:2017wau}
{\scshape DES} collaboration, T.~M.~C. Abbott et~al., \emph{{Dark Energy Survey
  Year 1 Results: Cosmological Constraints from Galaxy Clustering and Weak
  Lensing}},  \href{http://arxiv.org/abs/1708.01530}{{\tt 1708.01530}}.

\bibitem{Troxel:2017xyo}
{\scshape DES} collaboration, M.~A. Troxel et~al., \emph{{Dark Energy Survey
  Year 1 Results: Cosmological Constraints from Cosmic Shear}},
  \href{http://arxiv.org/abs/1708.01538}{{\tt 1708.01538}}.

\bibitem{Giusarma2016}
E.~Giusarma, M.~Gerbino, O.~Mena, S.~Vagnozzi, S.~Ho and K.~Freese,
  \emph{{Improvement of cosmological neutrino mass bounds}},
  \href{http://dx.doi.org/10.1103/PhysRevD.94.083522}{\emph{Phys. Rev.} {\bf
  D94} (2016) 083522}, [\href{http://arxiv.org/abs/1605.04320}{{\tt
  1605.04320}}].

\bibitem{Vagnozzi:2017ovm}
S.~Vagnozzi, E.~Giusarma, O.~Mena, K.~Freese, M.~Gerbino, S.~Ho et~al.,
  \emph{{Unveiling $\nu$ secrets with cosmological data: neutrino masses and
  mass hierarchy}},  \href{http://arxiv.org/abs/1701.08172}{{\tt 1701.08172}}.

\bibitem{Beutler:2014yhv}
{\scshape BOSS} collaboration, F.~Beutler et~al., \emph{{The clustering of
  galaxies in the SDSS-III Baryon Oscillation Spectroscopic Survey: signs of
  neutrino mass in current cosmological data sets}},
  \href{http://dx.doi.org/10.1093/mnras/stu1702}{\emph{Mon. Not. Roy. Astron.
  Soc.} {\bf 444} (2014) 3501--3516},
  [\href{http://arxiv.org/abs/1403.4599}{{\tt 1403.4599}}].

\bibitem{Brandbyge:2008js}
J.~Brandbyge and S.~Hannestad, \emph{{Grid Based Linear Neutrino Perturbations
  in Cosmological N-body Simulations}},
  \href{http://dx.doi.org/10.1088/1475-7516/2009/05/002}{\emph{JCAP} {\bf 0905}
  (2009) 002}, [\href{http://arxiv.org/abs/0812.3149}{{\tt 0812.3149}}].

\bibitem{Archidiacono:2015ota}
M.~Archidiacono and S.~Hannestad, \emph{{Efficient calculation of cosmological
  neutrino clustering in the non-linear regime}},
  \href{http://dx.doi.org/10.1088/1475-7516/2016/06/018}{\emph{JCAP} {\bf 1606}
  (2016) 018}, [\href{http://arxiv.org/abs/1510.02907}{{\tt 1510.02907}}].

\bibitem{Upadhye:2015lia}
A.~Upadhye, J.~Kwan, A.~Pope, K.~Heitmann, S.~Habib, H.~Finkel et~al.,
  \emph{{Redshift-space distortions in massive neutrino and evolving dark
  energy cosmologies}},
  \href{http://dx.doi.org/10.1103/PhysRevD.93.063515}{\emph{Phys. Rev.} {\bf
  D93} (2016) 063515}, [\href{http://arxiv.org/abs/1506.07526}{{\tt
  1506.07526}}].

\bibitem{AliHaimoud:2012vj}
Y.~Ali-Haimoud and S.~Bird, \emph{{An efficient implementation of massive
  neutrinos in non-linear structure formation simulations}},
  \href{http://dx.doi.org/10.1093/mnras/sts286}{\emph{Mon. Not. Roy. Astron.
  Soc.} {\bf 428} (2012) 3375--3389},
  [\href{http://arxiv.org/abs/1209.0461}{{\tt 1209.0461}}].

\bibitem{Shoji:2010hm}
M.~Shoji and E.~Komatsu, \emph{{Massive Neutrinos in Cosmology: Analytic
  Solutions and Fluid Approximation}},
  \href{http://dx.doi.org/10.1103/PhysRevD.81.123516,
  10.1103/PhysRevD.82.089901}{\emph{Phys. Rev.} {\bf D81} (2010) 123516},
  [\href{http://arxiv.org/abs/1003.0942}{{\tt 1003.0942}}].

\bibitem{Inman:2016qmg}
D.~Inman and U.-L. Pen, \emph{{Cosmic neutrinos: A dispersive and nonlinear
  fluid}}, \href{http://dx.doi.org/10.1103/PhysRevD.95.063535}{\emph{Phys.
  Rev.} {\bf D95} (2017) 063535}, [\href{http://arxiv.org/abs/1609.09469}{{\tt
  1609.09469}}].

\bibitem{Senatore:2017hyk}
L.~Senatore and M.~Zaldarriaga, \emph{{The Effective Field Theory of
  Large-Scale Structure in the presence of Massive Neutrinos}},
  \href{http://arxiv.org/abs/1707.04698}{{\tt 1707.04698}}.

\bibitem{Saito:2008bp}
S.~Saito, M.~Takada and A.~Taruya, \emph{{Impact of massive neutrinos on
  nonlinear matter power spectrum}},
  \href{http://dx.doi.org/10.1103/PhysRevLett.100.191301}{\emph{Phys. Rev.
  Lett.} {\bf 100} (2008) 191301}, [\href{http://arxiv.org/abs/0801.0607}{{\tt
  0801.0607}}].

\bibitem{Dakin:2017idt}
J.~Dakin, J.~Brandbyge, S.~Hannestad, T.~Haugbølle and T.~Tram,
  \emph{{$\nu$CO$N$CEPT: Cosmological neutrino simulations from the non-linear
  Boltzmann hierarchy}},  \href{http://arxiv.org/abs/1712.03944}{{\tt
  1712.03944}}.

\bibitem{Wright:2017dkw}
B.~S. Wright, H.~A. Winther and K.~Koyama, \emph{{COLA with massive
  neutrinos}},
  \href{http://dx.doi.org/10.1088/1475-7516/2017/10/054}{\emph{JCAP} {\bf 1710}
  (2017) 054}, [\href{http://arxiv.org/abs/1705.08165}{{\tt 1705.08165}}].

\bibitem{Viel:2010bn}
M.~Viel, M.~G. Haehnelt and V.~Springel, \emph{{The effect of neutrinos on the
  matter distribution as probed by the Intergalactic Medium}},
  \href{http://dx.doi.org/10.1088/1475-7516/2010/06/015}{\emph{JCAP} {\bf 1006}
  (2010) 015}, [\href{http://arxiv.org/abs/1003.2422}{{\tt 1003.2422}}].

\bibitem{Bird:2011rb}
S.~Bird, M.~Viel and M.~G. Haehnelt, \emph{{Massive Neutrinos and the
  Non-linear Matter Power Spectrum}},
  \href{http://dx.doi.org/10.1111/j.1365-2966.2011.20222.x}{\emph{Mon. Not.
  Roy. Astron. Soc.} {\bf 420} (2012) 2551--2561},
  [\href{http://arxiv.org/abs/1109.4416}{{\tt 1109.4416}}].

\bibitem{Brandbyge:2009ce}
J.~Brandbyge and S.~Hannestad, \emph{{Resolving Cosmic Neutrino Structure: A
  Hybrid Neutrino N-body Scheme}},
  \href{http://dx.doi.org/10.1088/1475-7516/2010/01/021}{\emph{JCAP} {\bf 1001}
  (2010) 021}, [\href{http://arxiv.org/abs/0908.1969}{{\tt 0908.1969}}].

\bibitem{Villaescusa-Navarro:2013pva}
F.~Villaescusa-Navarro, F.~Marulli, M.~Viel, E.~Branchini, E.~Castorina,
  E.~Sefusatti et~al., \emph{{Cosmology with massive neutrinos I: towards a
  realistic modeling of the relation between matter, haloes and galaxies}},
  \href{http://dx.doi.org/10.1088/1475-7516/2014/03/011}{\emph{JCAP} {\bf 1403}
  (2014) 011}, [\href{http://arxiv.org/abs/1311.0866}{{\tt 1311.0866}}].

\bibitem{Costanzi:2013bha}
M.~Costanzi, F.~Villaescusa-Navarro, M.~Viel, J.-Q. Xia, S.~Borgani,
  E.~Castorina et~al., \emph{{Cosmology with massive neutrinos III: the halo
  mass function andan application to galaxy clusters}},
  \href{http://dx.doi.org/10.1088/1475-7516/2013/12/012}{\emph{JCAP} {\bf 1312}
  (2013) 012}, [\href{http://arxiv.org/abs/1311.1514}{{\tt 1311.1514}}].

\bibitem{Castorina:2013wga}
E.~Castorina, E.~Sefusatti, R.~K. Sheth, F.~Villaescusa-Navarro and M.~Viel,
  \emph{{Cosmology with massive neutrinos II: on the universality of the halo
  mass function and bias}},
  \href{http://dx.doi.org/10.1088/1475-7516/2014/02/049}{\emph{JCAP} {\bf 1402}
  (2014) 049}, [\href{http://arxiv.org/abs/1311.1212}{{\tt 1311.1212}}].

\bibitem{Castorina2015}
E.~Castorina, C.~Carbone, J.~Bel, E.~Sefusatti and K.~Dolag, \emph{{DEMNUni:
  The clustering of large-scale structures in the presence of massive
  neutrinos}},
  \href{http://dx.doi.org/10.1088/1475-7516/2015/07/043}{\emph{JCAP} {\bf 1507}
  (2015) 043}, [\href{http://arxiv.org/abs/1505.07148}{{\tt 1505.07148}}].

\bibitem{Carbone:2016nzj}
C.~Carbone, M.~Petkova and K.~Dolag, \emph{{DEMNUni: ISW, Rees-Sciama, and
  weak-lensing in the presence of massive neutrinos}},
  \href{http://dx.doi.org/10.1088/1475-7516/2016/07/034}{\emph{JCAP} {\bf 1607}
  (2016) 034}, [\href{http://arxiv.org/abs/1605.02024}{{\tt 1605.02024}}].

\bibitem{Yu:2016yfe}
H.-R. Yu et~al., \emph{{Differential Neutrino Condensation onto Cosmic
  Structure}},  \href{http://arxiv.org/abs/1609.08968}{{\tt 1609.08968}}.

\bibitem{Emberson:2016ecv}
J.~D. Emberson et~al., \emph{{Cosmological neutrino simulations at extreme
  scale}}, \href{http://dx.doi.org/10.1088/1674-4527/17/8/85}{\emph{Res.
  Astron. Astrophys.} {\bf 17} (2017) 085},
  [\href{http://arxiv.org/abs/1611.01545}{{\tt 1611.01545}}].

\bibitem{RSD2017}
F.~Villaescusa-Navarro, A.~Banerjee, N.~Dalal, E.~Castorina, R.~Scoccimarro,
  R.~Angulo et~al., \emph{{The imprint of neutrinos on clustering in
  redshift-space}},  \href{http://arxiv.org/abs/1708.01154}{{\tt 1708.01154}}.

\bibitem{Adamek:2017uiq}
J.~Adamek, R.~Durrer and M.~Kunz, \emph{{Relativistic N-body simulations with
  massive neutrinos}},
  \href{http://dx.doi.org/10.1088/1475-7516/2017/11/004}{\emph{JCAP} {\bf 1711}
  (2017) 004}, [\href{http://arxiv.org/abs/1707.06938}{{\tt 1707.06938}}].

\bibitem{Inman:2015pfa}
D.~Inman, J.~D. Emberson, U.-L. Pen, A.~Farchi, H.-R. Yu and
  J.~Harnois-Déraps, \emph{{Precision reconstruction of the cold dark
  matter-neutrino relative velocity from $N$-body simulations}},
  \href{http://dx.doi.org/10.1103/PhysRevD.92.023502}{\emph{Phys. Rev.} {\bf
  D92} (2015) 023502}, [\href{http://arxiv.org/abs/1503.07480}{{\tt
  1503.07480}}].

\bibitem{DESI}
{\scshape DESI} collaboration, A.~Aghamousa et~al., \emph{{The DESI Experiment
  Part I: Science,Targeting, and Survey Design}},
  \href{http://arxiv.org/abs/1611.00036}{{\tt 1611.00036}}.

\bibitem{EUCLID}
R.~{Laureijs}, J.~{Amiaux}, S.~{Arduini}, J.~. {Augu{\`e}res}, J.~{Brinchmann},
  R.~{Cole} et~al., \emph{{Euclid Definition Study Report}}, {\emph{ArXiv
  e-prints} (Oct., 2011) }, [\href{http://arxiv.org/abs/1110.3193}{{\tt
  1110.3193}}].

\bibitem{WFIRST}
D.~{Spergel}, N.~{Gehrels}, J.~{Breckinridge}, M.~{Donahue}, A.~{Dressler},
  B.~S. {Gaudi} et~al., \emph{{Wide-Field InfraRed Survey
  Telescope-Astrophysics Focused Telescope Assets WFIRST-AFTA Final Report}},
  {\emph{ArXiv e-prints} (May, 2013) },
  [\href{http://arxiv.org/abs/1305.5422}{{\tt 1305.5422}}].

\bibitem{LSST}
{LSST Science Collaboration}, P.~A. {Abell}, J.~{Allison}, S.~F. {Anderson},
  J.~R. {Andrew}, J.~R.~P. {Angel} et~al., \emph{{LSST Science Book, Version
  2.0}}, {\emph{ArXiv e-prints} (Dec., 2009) },
  [\href{http://arxiv.org/abs/0912.0201}{{\tt 0912.0201}}].

\bibitem{Banerjee2016}
A.~Banerjee and N.~Dalal, \emph{{Simulating nonlinear cosmological structure
  formation with massive neutrinos}},
  \href{http://dx.doi.org/10.1088/1475-7516/2016/11/015}{\emph{JCAP} {\bf 1611}
  (2016) 015}, [\href{http://arxiv.org/abs/1606.06167}{{\tt 1606.06167}}].

\bibitem{1970PhFl...13.1819B}
J.~A. {Byers} and M.~{Grewal}, \emph{{Perpendicularly Propagating Plasma
  Cyclotron Instabilities Simulated with a One-Dimensional Computer Model}},
  \href{http://dx.doi.org/10.1063/1.1693160}{\emph{Physics of Fluids} {\bf 13}
  (July, 1970) 1819--1830}.

\bibitem{Ma:1993xs}
C.-P. Ma and E.~Bertschinger, \emph{{A Calculation of the full neutrino phase
  space in cold + hot dark matter models}},
  \href{http://dx.doi.org/10.1086/174298}{\emph{Astrophys. J.} {\bf 429} (1994)
  22}, [\href{http://arxiv.org/abs/astro-ph/9308006}{{\tt astro-ph/9308006}}].

\bibitem{healpix}
K.~M. Gorski, E.~Hivon, A.~J. Banday, B.~D. Wandelt, F.~K. Hansen, M.~Reinecke
  et~al., \emph{{HEALPix - A Framework for high resolution discretization, and
  fast analysis of data distributed on the sphere}},
  \href{http://dx.doi.org/10.1086/427976}{\emph{Astrophys. J.} {\bf 622} (2005)
  759--771}, [\href{http://arxiv.org/abs/astro-ph/0409513}{{\tt
  astro-ph/0409513}}].

\bibitem{2016arXiv160704590H}
D.~P. {Hardin}, T.~J. {Michaels} and E.~B. {Saff}, \emph{{A Comparison of
  Popular Point Configurations on $\mathbb{S}^2$}}, {\emph{ArXiv e-prints}
  (July, 2016) }, [\href{http://arxiv.org/abs/1607.04590}{{\tt 1607.04590}}].

\bibitem{Raccanelli_2017}
A.~{Raccanelli}, L.~{Verde} and F.~{Villaescusa-Navarro}, \emph{{Biases from
  neutrino bias: to worry or not to worry?}}, {\emph{ArXiv e-prints} (Apr.,
  2017) }, [\href{http://arxiv.org/abs/1704.07837}{{\tt 1704.07837}}].

\bibitem{Roncarelli_2017}
M.~{Roncarelli}, F.~{Villaescusa-Navarro} and M.~{Baldi}, \emph{{The kinematic
  Sunyaev-Zel'dovich effect of the large-scale structure (I): dependence on
  neutrino mass}}, \href{http://dx.doi.org/10.1093/mnras/stx170}{\emph{\mnras}
  {\bf 467} (May, 2017) 985--995}, [\href{http://arxiv.org/abs/1702.00676}{{\tt
  1702.00676}}].

\bibitem{Zennaro:2016nqo}
M.~Zennaro, J.~Bel, F.~Villaescusa-Navarro, C.~Carbone, E.~Sefusatti and
  L.~Guzzo, \emph{{Initial Conditions for Accurate N-Body Simulations of
  Massive Neutrino Cosmologies}},
  \href{http://dx.doi.org/10.1093/mnras/stw3340}{\emph{Mon. Not. Roy. Astron.
  Soc.} {\bf 466} (2017) 3244--3258},
  [\href{http://arxiv.org/abs/1605.05283}{{\tt 1605.05283}}].

\bibitem{Massara_2015}
E.~{Massara}, F.~{Villaescusa-Navarro}, M.~{Viel} and P.~M. {Sutter},
  \emph{{Voids in massive neutrino cosmologies}},
  \href{http://dx.doi.org/10.1088/1475-7516/2015/11/018}{\emph{\jcap} {\bf 11}
  (Nov., 2015) 018}, [\href{http://arxiv.org/abs/1506.03088}{{\tt
  1506.03088}}].

\bibitem{Springel:2005mi}
V.~Springel, \emph{{The Cosmological simulation code GADGET-2}},
  \href{http://dx.doi.org/10.1111/j.1365-2966.2005.09655.x}{\emph{Mon. Not.
  Roy. Astron. Soc.} {\bf 364} (2005) 1105--1134},
  [\href{http://arxiv.org/abs/astro-ph/0505010}{{\tt astro-ph/0505010}}].

\bibitem{CAMB}
A.~Lewis and A.~Challinor, ``{CAMB}.'' \url{http://camb.info/}.

\bibitem{CLASS1}
D.~{Blas}, J.~{Lesgourgues} and T.~{Tram}, \emph{{The Cosmic Linear Anisotropy
  Solving System (CLASS). Part II: Approximation schemes}},
  \href{http://dx.doi.org/10.1088/1475-7516/2011/07/034}{\emph{JCAP} {\bf 7}
  (July, 2011) 034}, [\href{http://arxiv.org/abs/1104.2933}{{\tt 1104.2933}}].

\bibitem{Massara_2014}
E.~{Massara}, F.~{Villaescusa-Navarro} and M.~{Viel}, \emph{{The halo model in
  a massive neutrino cosmology}},
  \href{http://dx.doi.org/10.1088/1475-7516/2014/12/053}{\emph{\jcap} {\bf 12}
  (Dec., 2014) 053}, [\href{http://arxiv.org/abs/1410.6813}{{\tt 1410.6813}}].

\bibitem{2017arXiv171001310C}
C.-T. {Chiang}, W.~{Hu}, Y.~{Li} and M.~{LoVerde}, \emph{{Scale-dependent bias
  and bispectrum in neutrino separate universe simulations}}, {\emph{ArXiv
  e-prints} (Oct., 2017) }, [\href{http://arxiv.org/abs/1710.01310}{{\tt
  1710.01310}}].

\bibitem{Smith:2002dz}
{\scshape VIRGO Consortium} collaboration, R.~E. Smith, J.~A. Peacock,
  A.~Jenkins, S.~D.~M. White, C.~S. Frenk, F.~R. Pearce et~al., \emph{{Stable
  clustering, the halo model and nonlinear cosmological power spectra}},
  \href{http://dx.doi.org/10.1046/j.1365-8711.2003.06503.x}{\emph{Mon. Not.
  Roy. Astron. Soc.} {\bf 341} (2003) 1311},
  [\href{http://arxiv.org/abs/astro-ph/0207664}{{\tt astro-ph/0207664}}].

\bibitem{Takahashi:2012em}
R.~Takahashi, M.~Sato, T.~Nishimichi, A.~Taruya and M.~Oguri, \emph{{Revising
  the Halofit Model for the Nonlinear Matter Power Spectrum}},
  \href{http://dx.doi.org/10.1088/0004-637X/761/2/152}{\emph{Astrophys. J.}
  {\bf 761} (2012) 152}, [\href{http://arxiv.org/abs/1208.2701}{{\tt
  1208.2701}}].

\bibitem{Lawrence2017}
E.~Lawrence, K.~Heitmann, J.~Kwan, A.~Upadhye, D.~Bingham, S.~Habib et~al.,
  \emph{{The Mira-Titan Universe II: Matter Power Spectrum Emulation}},
  \href{http://dx.doi.org/10.3847/1538-4357/aa86a9}{\emph{Astrophys. J.} {\bf
  847} (2017) 50}, [\href{http://arxiv.org/abs/1705.03388}{{\tt 1705.03388}}].

\bibitem{DeBernardis:2009di}
F.~De~Bernardis, T.~D. Kitching, A.~Heavens and A.~Melchiorri,
  \emph{{Determining the Neutrino Mass Hierarchy with Cosmology}},
  \href{http://dx.doi.org/10.1103/PhysRevD.80.123509}{\emph{Phys. Rev.} {\bf
  D80} (2009) 123509}, [\href{http://arxiv.org/abs/0907.1917}{{\tt
  0907.1917}}].

\bibitem{Jimenez:2010ev}
R.~Jimenez, T.~Kitching, C.~Pena-Garay and L.~Verde, \emph{{Can we measure the
  neutrino mass hierarchy in the sky?}},
  \href{http://dx.doi.org/10.1088/1475-7516/2010/05/035}{\emph{JCAP} {\bf 1005}
  (2010) 035}, [\href{http://arxiv.org/abs/1003.5918}{{\tt 1003.5918}}].

\bibitem{Abel2011}
T.~Abel, O.~Hahn and R.~Kaehler, \emph{{Tracing the Dark Matter Sheet in Phase
  Space}}, \href{http://dx.doi.org/10.1111/j.1365-2966.2012.21754.x}{\emph{Mon.
  Not. Roy. Astron. Soc.} {\bf 427} (2012) 61--76},
  [\href{http://arxiv.org/abs/1111.3944}{{\tt 1111.3944}}].

\bibitem{powell2015a}
D.~Powell and T.~Abel, \emph{An exact general remeshing scheme applied to
  physically conservative voxelization}, {\emph{Journal of Computational
  Physics} {\bf 297} (2015) 340--356}.

\bibitem{powell2015b}
D.~M. Powell, \emph{{r3d}: Software for fast, robust geometric operations in
  {3D} and {2D}}, {\emph{LA-UR-15-26964} (2015) }.

\bibitem{wojtak2016}
R.~{Wojtak}, D.~{Powell} and T.~{Abel}, \emph{{Voids in cosmological
  simulations over cosmic time}},
  \href{http://dx.doi.org/10.1093/mnras/stw615}{\emph{\mnras} {\bf 458} (June,
  2016) 4431--4442}, [\href{http://arxiv.org/abs/1602.08541}{{\tt
  1602.08541}}].

\bibitem{sousbie2015}
T.~{Sousbie} and S.~{Colombi}, \emph{{ColDICE: a parallel Vlasov-Poisson solver
  using moving adaptive simplicial tessellation}}, {\emph{ArXiv e-prints}
  (Sept., 2015) }, [\href{http://arxiv.org/abs/1509.07720}{{\tt 1509.07720}}].

\bibitem{hahn2016}
O.~{Hahn} and R.~E. {Angulo}, \emph{{An adaptively refined phase-space element
  method for cosmological simulations and collisionless dynamics}},
  \href{http://dx.doi.org/10.1093/mnras/stv2304}{\emph{\mnras} {\bf 455} (Jan.,
  2016) 1115--1133}, [\href{http://arxiv.org/abs/1501.01959}{{\tt
  1501.01959}}].

\bibitem{kh2016}
J.~Kates-Harbeck, S.~Totorica, J.~Zrake and T.~Abel, \emph{{Simplex-in-cell
  technique for collisionless plasma simulations}},
  \href{http://dx.doi.org/10.1016/j.jcp.2015.10.017}{\emph{J. Comput. Phys.}
  {\bf 304} (2016) 231--251}.

\bibitem{2013JCAP...03..019V}
F.~{Villaescusa-Navarro}, S.~{Bird}, C.~{Pe{\~n}a-Garay} and M.~{Viel},
  \emph{{Non-linear evolution of the cosmic neutrino background}},
  \href{http://dx.doi.org/10.1088/1475-7516/2013/03/019}{\emph{\jcap} {\bf 3}
  (Mar., 2013) 019}, [\href{http://arxiv.org/abs/1212.4855}{{\tt 1212.4855}}].

\bibitem{Betts:2013uya}
S.~Betts et~al., \emph{{Development of a Relic Neutrino Detection Experiment at
  PTOLEMY: Princeton Tritium Observatory for Light, Early-Universe,
  Massive-Neutrino Yield}},  in \emph{{Proceedings, 2013 Community Summer Study
  on the Future of U.S. Particle Physics: Snowmass on the Mississippi
  (CSS2013): Minneapolis, MN, USA, July 29-August 6, 2013}}, 2013.
\newblock \href{http://arxiv.org/abs/1307.4738}{{\tt 1307.4738}}.

\bibitem{2017ascl.soft06006B}
S.~{Bird}, ``{GenPK: Power spectrum generator}.'' Astrophysics Source Code
  Library, June, 2017.

\bibitem{2014ascl.soft11001T}
R.~{Thompson}, ``{pyGadgetReader: GADGET snapshot reader for python}.''
  Astrophysics Source Code Library, Nov., 2014.

\end{thebibliography}\endgroup
\bibliographystyle{JHEP}
\end{document}